\documentclass[prb,twocolumn,showpacs,superscriptaddress]{revtex4}

\usepackage{graphicx}
\usepackage{amsmath}
\usepackage{amssymb}

\newcommand{\be}{\begin{equation}}
\newcommand{\ee}{\end{equation}}
\newcommand{\ba}{\begin{eqnarray}}
\newcommand{\ea}{\end{eqnarray}}

\newcount\bozza \bozza=0
\ifnum\bozza=1
\newdimen\shift \shift=-2truecm
\def\lb#1{%
{\label{#1}\rlap{\kern\shift{$\scriptstyle#1$}}}}
\else\def\lb#1{\label{#1}} \fi

\begin{document}


\title{Density of states of relativistic and nonrelativistic two-dimensional electron gases
in  a uniform magnetic and  Aharonov-Bohm fields}

\author{A.O. Slobodeniuk}
\email{aslobodeniuk@gmail.com}
\affiliation{Bogolyubov Institute for Theoretical Physics, National Academy of Science of Ukraine, 14-b
        Metrologicheskaya Street, Kiev, 03680, Ukraine}

\author{S.G.~Sharapov}
\email{sharapov@bitp.kiev.ua}
\affiliation{Bogolyubov Institute for Theoretical Physics, National Academy of Science of Ukraine, 14-b
        Metrologicheskaya Street, Kiev, 03680, Ukraine}


\author{V.M.~Loktev}
\email{vloktev@bitp.kiev.ua}
\affiliation{Bogolyubov Institute for Theoretical Physics, National Academy of Science of Ukraine, 14-b
        Metrologicheskaya Street, Kiev, 03680, Ukraine}
\affiliation{National Technical University of Ukraine "KPI",
   37 Peremogy Ave., Kiev  03056, Ukraine}

\date{\today }

\begin{abstract}
We study the electronic properties of 2D electron gas (2DEG) with quadratic dispersion and
with relativistic dispersion as in graphene  in the inhomogeneous magnetic field consisting of
the Aharonov-Bohm flux and a constant background field. The total and local density of states (LDOS)
are obtained on the base of the analytic solutions of the Schr\"{o}dinger and Dirac equations in the inhomogeneous
magnetic field. It is shown that as it was in the situation with a pure Aharonov-Bohm flux,
in the case of graphene there is an excess of LDOS near the vortex, while in 2DEG the LDOS is depleted.
This results in excess of the induced by the vortex DOS in  graphene  and in its depletion in 2DEG.
\end{abstract}

\pacs{03.65.-w, 73.20.At, 72.10.Fk}


\maketitle

\section{Introduction}
\label{sec:intro}

The continuous linear energy dispersion $E(\mathbf{k}) = \pm \hbar v_F |\mathbf{k}|$ of
the Dirac quasiparticle excitations  when the homogeneous magnetic field $B$ is applied perpendicular
to its two-dimensional (2D) plane transforms into the discrete Landau levels (LLs)
\be
\label{Dirac-LL}
E_n = \pm \epsilon_0 \sqrt{2 n}, \quad n=0,1,2\ldots,
\ee
observed in graphene.
Here $\mathbf{k}$ is the momentum measured from $\mathbf{K}_{\pm}$ points,
$\epsilon_0 = \sqrt{\hbar v_F^2 eB/c}$ is the relativistic Landau scale with
$v_F$ being the Fermi velocity.
The spectrum (\ref{Dirac-LL}) is characteristic of Dirac fermions  and
the breakthrough in experimental studies of graphene is caused not only by its fabrication
\cite{Novoselov2004Science}, but also by the demonstration of its unique electronic properties
\cite{Geim2005Nature,Kim2005Nature} that follow from the unusual spectrum (\ref{Dirac-LL}).

The hallmark of this spectrum is the zero energy field independent lowest LL
whose existence does not in fact depend on the homogeneity of the field.\cite{Thaller:book}
In general the inhomogeneous magnetic perturbation can be presented as a sum of a constant
(averaged over the system) field and a field localized in some regions of the 2D system.
A limiting case of the perturbation can be presented by the Aharonov-Bohm 
field which is created by an infinitely long and infinitesimally thin solenoid.

The purpose of the present paper is to study the electronic excitations in graphene
in a field consisting of the Aharonov-Bohm flux and a constant background magnetic field.
As in the first publication,\cite{Slobodeniuk2010PRB} where we studied the  Aharonov-Bohm flux only,
our main goal is the investigation of the local density of states
(LDOS). We find that the demonstrated in Ref.~\onlinecite{Slobodeniuk2010PRB}
rather peculiar behavior of LDOS in Dirac theory with Aharonov-Bohm field  persists in the
presence of the constant background field. We expect that this behavior can be observed
in scanning tunneling spectroscopy  measurements for graphene penetrated by vortices
from a type-II superconductor on top of it.  We also compare the obtained expressions with the corresponding
results for two-dimensional electron gas (2DEG) with a quadratic dispersion, where the singular behavior of
the LDOS is absent.

In practice, such a magnetic field configuration may be obtained when a type-II superconductor
is placed on top of graphene. In the previous publication, we considered an idealized picture
when the vortex is single and there is no impact from other Abrikosov vortices. Now the constant
background field is supposed to mimic the impact of the other vortices penetrating graphene.
It is worth to stress that devices like this, with a superconducting film grown on top of a
semiconducting heterojunction (such as GaAs/AlGaAs) hosting a 2DEG, have in fact been fabricated
twenty years ago,\cite{Bending1990PRL,Geim1992PRL} so it should be possible to fabricate the
graphene based devices. While normally the 2DEG is buried deep in a semiconducting heterostructure which
makes the LDOS measurements problematic,\cite{Hashimoto2008PRL} the graphene surface is open to
the LDOS measurements. While initially the STS measurements were done on graphene flakes on graphite\cite{Li2009PRL},
recently these measurements were carried out on exfoliated graphene samples deposited on a chlorinated SiO$_2$
thermal oxide tuning the density through the Si backgate.\cite{Luican2011PRB}
So far all these measurements were done in  a homogeneous magnetic field and showed a single sequence of
pronounced LL peaks corresponding to massless Dirac fermions expected of pristine graphene.

In a wider context, the inhomogeneous vortex-like field configurations arise due the topological
defects in graphene that result in the pseudomagnetic field vortices, see, e.g.,
Refs.~\onlinecite{Cortijo2007NPB,Sitenko2007NPB,Vozmediano2010PR}. Interestingly, even the nonsingular
pseudomagnetic field configuration created by a curved bump on flat graphene\cite{Juan2007PRB} results
in the oscillations of the LDOS similar to the long-distance behavior of LDOS induced by
the Abrikosov's vortex.\cite{Slobodeniuk2010PRB} As proven experimentally,
a strong localized pseudomagnetic field can be induced in graphene by a strain and the
corresponding LLs are observed in the STS measurements.\cite{Levy2010Science}

Thus we hope that the
combination of the  vortex $+$ constant background field considered in the present paper should be useful
not only for the studies that involve a real magnetic field, but also for the problems that involve
the superposition of magnetic and pseudomagnetic fields.

The paper is organized as follows. In Sec.~\ref{sec:model} we introduce the model
Hamiltonians and discuss the configuration of the magnetic field
and the regularization of the Aharonov-Bohm potential used in this work.
Sec.~\ref{sec:Schrodinger} is devoted to the nonrelativistic case, and the relativistic case
is discussed in detail in Sec.~\ref{sec:Dirac}.
The structure of both sections is the same: we consider the solutions of the corresponding
Schr\"{o}dinger or Dirac equation, which allow to write down a general representation for the LDOS in
Secs.~\ref{sec:nonrel-solutions} and \ref{sec:Dirac-solutions}. Then, a more simple analysis of
the DOS is made in  Secs.~\ref{sec:DOS-nonrel} and \ref{sec:DOS-rel}, while
the behavior of the LDOS is studied in Secs.~\ref{sec:LDOS-nonrel} and \ref{sec:LDOS-rel}.
In Sec.~\ref{sec:concl}, our final results are summarized.
The method of the calculation of the LDOS is explained in Appendix~\ref{sec:A}, where as an example we
firstly calculate the LDOS  in a constant magnetic field for the nonrelativistic case.
Since the problem with Aharonov-Bohm vortex has to be treated in the symmetric gauge, the calculation of
the LDOS in Appendix~\ref{sec:A} involves the sum over the azimuthal quantum number, which is calculated
in Appendix~\ref{sec:B}. The full DOS is calculated in Appendix~\ref{sec:C}. The LDOS, both in nonrelativistic
and relativistic cases, is expressed in terms of the function calculated in Appendix~\ref{sec:D}. The Dirac
equation in the magnetic field consisting of the Aharonov-Bohm flux and a constant background field
is solved in Appendix~\ref{sec:E}.

\section{Models and main notations}
\label{sec:model}

As in Ref.~\onlinecite{Slobodeniuk2010PRB}, we consider both nonrelativistic and relativistic Hamiltonians.
The 2D nonrelativistic (Schr\"{o}dinger) Hamiltonian has the standard form
\begin{equation}
\label{Hamilton-nonrel}
H_S = -\frac{\hbar^2}{2M} (D_1^2 +D_2^2),
\end{equation}
where $D_j = \nabla_j + i e/\hbar c A_j $, $j=1,2$, with the vector potential $\mathbf{A}$,
Planck's constant $\hbar$, and the velocity of light $c$
describes a spinless particle with a mass $M$ and charge $-e < 0$.

The Dirac quasiparticle in graphene is described by the Hamiltonian
\begin{equation}
\label{Hamilton-rel}
H_D= - i \hbar v_F \beta (\gamma_1 D_1 + \gamma_2 D_2) + \Delta \beta,
\end{equation}
where the matrices $\beta$ and $\beta \gamma_j$ are defined in terms of the Pauli
matrices as
\begin{equation}
\label{matrices}
\beta = \sigma_3, \qquad \beta \gamma_j = (\sigma_1, \zeta \sigma_2).
\end{equation}
Here $\zeta = \pm 1$ labels two unitary inequivalent representations of $2\times2$ gamma matrices in $2+1$ dimension,
so that one considers a pair of Dirac equations corresponding to two inequivalent
$\mathbf{K}_{\pm}$ points of graphene's Brillouin zone.  The
spin degree of freedom is not included neither in Eq.~(\ref{Hamilton-nonrel}) nor in Eq.~(\ref{Hamilton-rel}).
In Eq.~(\ref{Hamilton-rel}), $v_F$ is the Fermi velocity and $\Delta$ is the Dirac mass (or gap).
An overview of its physical origin is given in\cite{Slobodeniuk2010PRB}
(see also a review \cite{Gusynin2007IJMPB}). Here we only point out that the presence of
a finite $\Delta$ allows one to distinguish unambiguously positive and negative energy solutions.

There are numerous studies of the Dirac fermions in the field of a singular Aharonov-Bohm vortex
(see, e.g., Refs.~\onlinecite{Gerbert1989PRD,Jackiw1991book,Sitenko1996PLB}) and, in particular,
of this vortex and a uniform magnetic field\cite{Falomir2001JPA,Gavrilov2003EPJC} devoted to the mathematical
aspects of the problem  such as self-adjoint extension of the Dirac operator.
As in the previous article to avoid the mathematical difficulties related to a singular nature of the Aharonov-Bohm
potential at the origin, we consider a regularized potential\cite{Alford1989NPB,Hagen1990PRL} that depends on
the dimensional parameter $R$:
\begin{equation}\label{reg-potential}
\mathbf{A}(\mathbf{r})=A_\varphi(r)\mathbf{e}_\varphi, \qquad
A_\varphi(r)=\frac{Br}{2}+\frac{\Phi_0\eta}{2 \pi r}\theta(r-R),
\end{equation}
where $\mathbf{r} = (r, \varphi,z)$,
$\Phi_0 \eta$ is the flux of the vortex expressed via magnetic flux quantum of the electron
$\Phi_0 = h c/e$ with $\eta \in [0,1[$. The value $\eta = 1/2$ corresponds to the
Abrikosov's vortex flux.  The corresponding magnetic field
\begin{equation}
\label{magn-field}
\mathbf{B}(\mathbf{r}) = \nabla \times \mathbf{A} = \left(B + \frac{\eta \Phi_0}{2 \pi R} \delta(r-R) \right) \mathbf{e}_z.
\end{equation}
The radius $R$  of the flux tube determines the region $r >R$ where the regularized potential coincides
with the potential of the problem with Aharonov-Bohm potential, while for $r < R$ it describes
a particle moving in a constant magnetic field. The solution of
the problem is found by matching the solutions obtained in these regions.
The limit $R \to 0$ can be taken at the end and allows to avoid the formal complications.
As was shown in Ref.~\onlinecite{Hagen1990PRL}, the final answer does not depend on the specific form of the
regularizing potential provided that the profile of the magnetic field is nonsingular at the origin.

We also mention recent works\cite{Jackiw2009PRB,Milstein2010} where the induced by the Aharonov-Bohm
field charge density and current were studied for the massless Dirac fermions.
In the first paper,\cite{Jackiw2009PRB} an infinitesimally thin solenoid was considered. The
regularization by a magnetic flux tube of a small radius $R$ as in the present work is considered
in the second paper.\cite{Milstein2010} It is shown that in the limit $R \to 0$ the induced current is
a periodic function of the magnetic flux irrespectively of the magnetic field distribution inside
the flux tube and whether the region inside the flux tube is forbidden or not for penetration by electrons.
Also the value of the self-adjoint extension parameter is fixed by the regularization.
The properties of the quasibound states in 2DEG with parabolic dispersion as well as Dirac electrons with
linear dispersion in the presence of a circular step magnetic field profile were recently studied in
Ref.~\onlinecite{Masir2009PRB}.

\section{Nonrelativistic  case}
\label{sec:Schrodinger}

In this section, we consider the solutions of the Schr\"{o}dinger equation
\begin{equation}
\label{nonrel-eq}
H_S \psi(\mathbf{r}) = E \psi(\mathbf{r})
\end{equation}
in polar coordinates $\mathbf{r}=(r,\varphi)$ and using them we obtain the full and local DOS.
These results are important not only for comparison with the relativistic case, but
also because the relativistic result is constructed using the nonrelativistic one.

\subsection{Solution of the Schr\"{o}dinger equation and
general representation for the local density of states and its limiting $\eta=0$ case}
\label{sec:nonrel-solutions}

Technically, to obtain the solutions of Eq.~(\ref{nonrel-eq}) in the regularized potential (\ref{reg-potential}),
one should solve this equation in two regions: $r<R$ and $r>R$.
Since in the first domain, $r <R$, the potential is nonsingular, only a regular in the limit $r \to 0$ solution of the radial differential equation is admissible. In the second domain, $r>R$, the solution contains both regular and singular in
the limit $r \to 0$ terms. The values of the
relative weights of them can be found by matching radial components and their derivatives
at $r = R$. Finally, it turns out that in the limit $R \to 0$ only the regular solution survives and the wave
function takes the form
\begin{equation}
\label{solution-nonrel}
\psi_{n,m}(r,\varphi)=A_{n,m} e^{im\varphi} y^{|m+\eta|/2}e^{-y/2}L_n^{|m+\eta|}(y),
\end{equation}
which also follows from the Schr\"{o}dinger equation with a singular vortex.
Here, the dimensionless variable $y \equiv r^2/(2 l^2)$ is expressed via the magnetic length
$l= (\hbar c/eB)^{1/2}$, $L_n^{\alpha}(y)$ is the generalized Laguerre polynomial and
the normalization constant $A_{n,m}$ is given by
\begin{equation}
 A_{n,m}^2=\frac{n!}{2\pi l^2\Gamma(n+|m+\eta|+1)}.
\end{equation}
 The corresponding to the wave function (\ref{solution-nonrel}) eigenenergy  is equal to
\begin{equation}
\label{spectrum-nonrel}
E_{n,m} = \frac{\hbar \omega_c}{2}(2n+1 + |m+ \eta| + m + \eta ),
\end{equation}
where the cyclotron frequency $\omega_c = e B/Mc$, the radial quantum number  $n=0,1,\ldots$, and
the azimuthal quantum number $m=-\infty,\ldots,-1,0,1, \ldots, \infty$.
In what follows, it is convenient to express all energies of the nonrelativistic
problem in terms of the energy $E_0 \equiv \hbar \omega_c/2$.

Having the wave function, one can calculate the LDOS using the representation
\begin{equation}
\label{LDOS-def}
N(\mathbf{r},E,B)= \sum_{n,m} |\psi_{n,m} (\mathbf{r})|^2\delta(E-E_{n,m}).
\end{equation}
In contrast to the previous article,\cite{Slobodeniuk2010PRB} the
presence of a constant magnetic field makes all energy spectra  discrete, which demands
some regularization of the $\delta$ function in Eq.~(\ref{LDOS-def}). For this purpose,
we introduce a widening of the LLs to a Lorentzian shape:
\begin{equation}
\label{LL2Lorentzian}
 \delta(E-E_{n,m})\rightarrow\frac{1}{\pi} \mbox{Im} \frac{1}{E_{n,m}-E-i \Gamma},
\end{equation}
where $\Gamma$ is the LL width. Such a simple broadening of LLs
with a constant $\Gamma$ was found to be a rather good approximation valid in not
very strong magnetic fields.\cite{Shoenberg:book}

\begin{figure}[h]
\centering{
\includegraphics[width=0.5\textwidth]{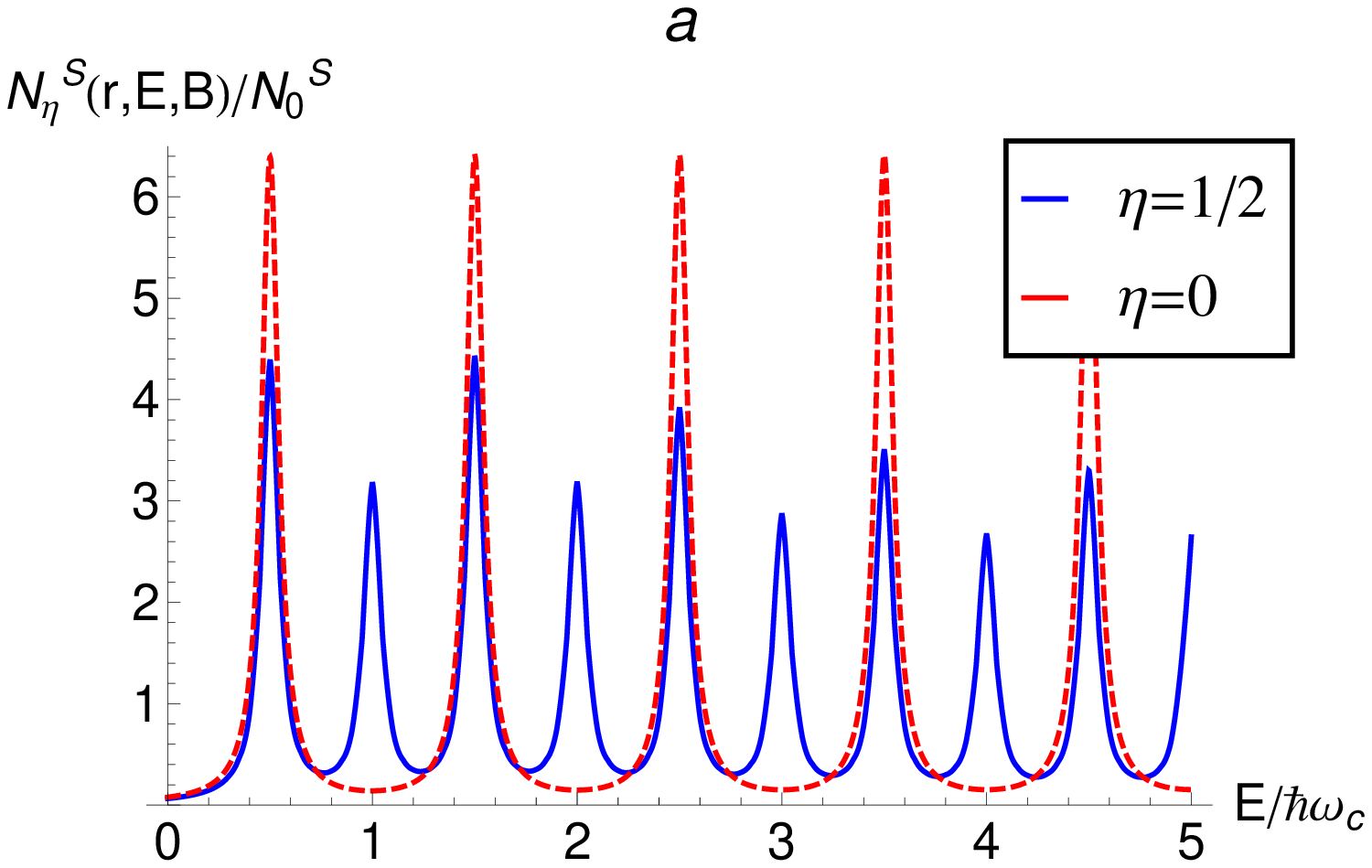}}
\centering{
\includegraphics[width=0.5\textwidth]{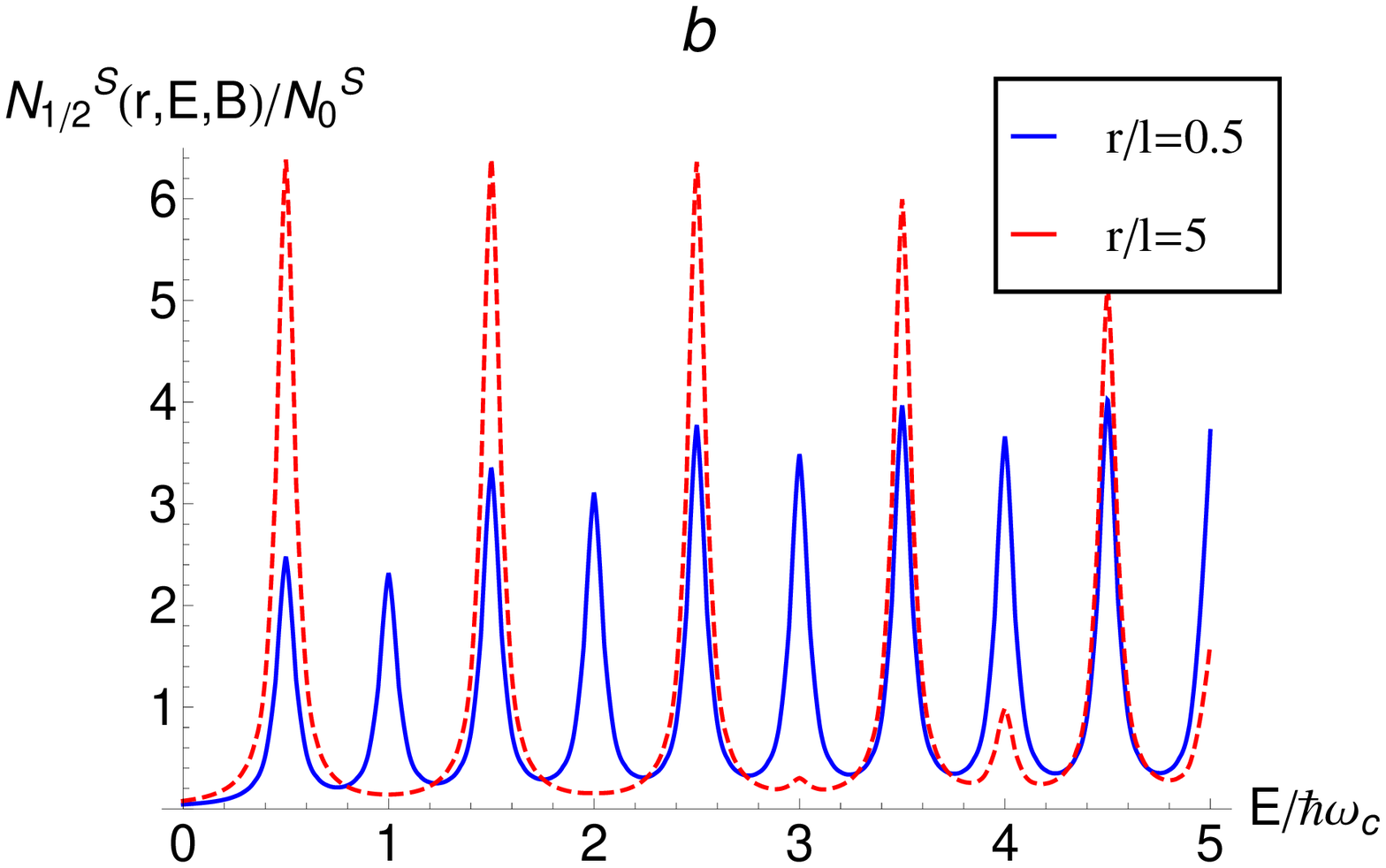}}
\caption{(Color online)
The normalized full LDOS  $N_{\eta}^{\mathrm{S}}(r,E,B)/N_0^{\mathrm{S}}$ as a function of
$E$ in the units of $\hbar \omega_c$. (a) $\eta = 0$ (no vortex and LDOS is $r$ independent)
and $\eta = 1/2$ for $r = l$. (b) Both lines are for $\eta = 1/2$, $r =0.5 l$, and $r = 5l$.
In all cases, the width is $\Gamma = 0.05 \hbar \omega_c$.}
\label{fig:1}
\end{figure}

To illustrate the method of calculation  in Appendix~\ref{sec:A}, we derive the LDOS for the simplest case ($\eta=0$)
of the constant magnetic field without vortex
\begin{equation}
\label{LDOS-nonrel-const}
N^{\mathrm{S}}_0(E,B)=  -\frac{N_0^{\mathrm{S}}}{\pi} \mbox{Im} \psi\left(\frac{1}{2} - \frac{E+i \Gamma}{\hbar \omega_c} \right).
\end{equation}
Here, $N_0^{\mathrm{S}} = M/(2 \pi \hbar^2)$ is a free DOS of 2DEG per spin and unit area and we omitted the
$\mathbf{r}$ dependence of the LDOS, because it is absent in the homogeneous field.
One can readily obtain Eq.~(\ref{LDOS-nonrel-const}) in a much simplier way\cite{Sharapov2004PRB}
starting from the usual Landau spectrum
\begin{equation}
\label{Schrodinger-LL}
E_n = \hbar \omega_c \left( n+ \frac{1}{2} \right),
\end{equation}
which follows from the spectrum (\ref{spectrum-nonrel})
for $\eta=0$, when one relabels $n + (|m|+m)/2 \to n$. Here, the relabeled $n$ corresponds to the LL index
rather than the radial quantum number. Nevertheless, in Appendix~\ref{sec:A} we proceeded from Eq.~(\ref{spectrum-nonrel})
to illustrate how to deal with a spectrum that is also dependent on the azimuthal quantum number $m$.
As seen in Fig.~\ref{fig:1}~(a) on the dashed (red) curve, Eq.~(\ref{LDOS-nonrel-const}) describes the usual
quantum magnetic oscillations of the DOS.
One can extract them analytically using the reflection formula (\ref{psi-inv}). Integrating the DOS
over the energy with the thermal factor $\ln \left[1+ \exp\left(\frac{\mu - E}{T}\right) \right]$,
one can obtain the thermodynamic potential, whose derivative with respect to the magnetic field gives magnetization.
The corresponding oscillations of the magnetization are known as the de Haas-van Alphen effect.\cite{Shoenberg:book}

In a similar fashion we obtain in Appendix~\ref{sec:A} the expression for the LDOS perturbation,
$\Delta N_\eta^{\mathrm{S}}(\mathbf{r},E,B) = N_\eta^{\mathrm{S}}(\mathbf{r},E,B) -  N_0^{\mathrm{S}}(\mathbf{r},E)$ induced by the vortex
\begin{widetext}
\begin{equation}
\label{delta1-nonrel-LDOS}
\Delta N_\eta^{\mathrm{S}}(\mathbf{r},E,B) = -\frac{M}{(\pi\hbar)^2}\frac{\sin\pi\eta}{\pi}
\mbox{Im} \left[ \int_0^\infty  d\beta e^{-(\delta+\beta)}e^{-\beta z}
\frac{e^{-y\coth(\delta+\beta)}}{1-e^{-2(\delta+\beta)}}
\int_{-\infty}^\infty  d\omega e^{-y\cosh\omega/\sinh(\delta+\beta)}
\frac{e^{-\eta(\delta+\beta+\omega)}}{1+e^{-(\delta+\beta+\omega)}}\right].
 \end{equation}
\end{widetext}
Here, $N_\eta^{\mathrm{S}}(\mathbf{r},E,B)$ is the LDOS in the presence of the constant field and vortex and $N^{\mathrm{S}}_0(\mathbf{r},E,B)$
is the LDOS in the constant magnetic field without vortex (the argument $\mathbf{r}$ is present to distinguish the LDOS from the DOS). This expression has to be calculated for $z>0$ with the analytic continuation $z \to - (E+ i\Gamma)/E_0$
done at the end of the calculation.
The representation (\ref{delta1-nonrel-LDOS}) for the LDOS is our starting point for the analysis of the LDOS and DOS.
Next, in  Sec.~\ref{sec:DOS-nonrel} we begin with a simpler case of the DOS and return to the LDOS in Sec.~\ref{sec:LDOS-nonrel}.

\subsection{The density of states}
\label{sec:DOS-nonrel}

While in the constant magnetic field  the LDOS is position independent and is related to the
full DOS by the 2D volume (area) of the system factor $V_{\mathrm{2D}}$, this is not so in the presence of the vortex
when the LDOS is position dependent. Then the full DOS per spin projection is obtained from the LDOS (\ref{LDOS-def})
by integrating over the space coordinates
\begin{equation}
\label{DOS-nonrel}
N_\eta(E,B)= \int_0^{2\pi} d \varphi \int_0^\infty r d r N_\eta(\mathbf{r},E,B).
\end{equation}
The details of the derivation of the DOS difference,
$\Delta N_\eta^{\mathrm{S}}(E,B)= N_\eta^{\mathrm{S}}(E,B) - N_0^{\mathrm{S}}(E,B)$ with $N_0^{\mathrm{S}}(E)$ being
the full DOS in the presence of the constant field without vortex, are given in Appendix~\ref{sec:C}. We obtain
\begin{equation}
\label{DOS-nonrel-vortex-const}
\begin{split}
&\Delta N_\eta^{\mathrm{S}}(E,B)= \frac{1}{\pi \hbar \omega_c}
\mbox{Im}\left\{ \left(\frac{1}{2}+\frac{E+i \Gamma}{\hbar \omega_c} - \eta \right) \right. \\
& \times \left.  \left[ \psi\left(\frac{1}{2} - \frac{E+i \Gamma}{\hbar \omega_c}\right) -
\psi\left(\frac{1}{2} - \frac{E+i \Gamma}{\hbar \omega_c} + \eta\right) \right] \right\}.
\end{split}
\end{equation}
Since the digamma function $\psi(z)$ has simple poles for $z=0,-1,-2,\ldots$, it is easy to see in the clean limit $\Gamma\to0$,
that the DOS difference (\ref{DOS-nonrel-vortex-const}) reduces to a set of $\delta$ peaks corresponding to the LLs:
\begin{equation}
\label{DOS-nonrel-vortex-const-delta}
\begin{split}
\Delta N_\eta^{\mathrm{S}}(E,B)= &-\sum_{n=0}^{\infty} (n+1 - \eta) \delta \left(E -\hbar \omega_c \left(n+ \frac{1}{2} \right) \right)\\
&+ \sum_{n=0}^{\infty} (n+1)\delta \left(E -\hbar \omega_c \left(n+ \frac{1}{2} + \eta \right) \right)
\end{split}
\end{equation}
The physical meaning of (\ref{DOS-nonrel-vortex-const-delta}) is that\cite{Desbois1997NPB} on each LL
$E_{n} = \hbar \omega_c (n+1/2)$, $n+1 - \eta$ states disappear and $n+1$ appear at the energy $E_{n} =
\hbar \omega_c (n+1/2+\eta)$.

The limit of zero field, $B\to 0$, can be obtained from Eq.~(\ref{DOS-nonrel-vortex-const})
using the asymptotic expansion
\begin{equation}
\label{psi-expand}
\psi(z)= \ln z  -
\frac{1}{2\,z} - \frac{1}{12\,z^2} + O\left(\frac{1}{z^4}\right).
\end{equation}
Then  in the limit $\Gamma \to 0$, we
reproduce the Aharonov-Bohm depletion of the DOS\cite{Desbois1997NPB,Moroz1996PRA,Slobodeniuk2010PRB}
at the bottom of the spectrum
\begin{equation}
\label{DOS-nonrel-depletion}
\begin{split}
\Delta N_\eta^{\mathrm{S}}(E,B=0) &= N_\eta^{\mathrm{S}}(E,B=0)- V_{2D} N_0^{\mathrm{S}} = \\
& = -\frac{1}{2} \eta (1-\eta) \delta(E)
\end{split}
\end{equation}
caused by an isolated vortex. Integrating Eqs.~(\ref{DOS-nonrel-depletion}) and (\ref{DOS-nonrel-vortex-const-delta})
(with an appropriate regularization)
one can check that the total deficit of the states induced by the vortex
\begin{equation}
\label{deficit}
\Delta N_\eta^{\mathrm{S}} \equiv \int_{-\infty}^{\infty} d E \Delta N_\eta^{\mathrm{S}}(E,B) = -\frac{1}{2} \eta (1-\eta)
\end{equation}
does not depend on the strength $B$ of the nonsingular background field.

\subsection{The local density of states}
\label{sec:LDOS-nonrel}

The regularization parameter $\delta$  in Eq.~(\ref{delta1-nonrel-LDOS}) is important for the calculation of
the DOS made in Appendix~\ref{sec:B}, the integrand of   Eq.~(\ref{delta1-nonrel-LDOS}) remains regular
even in the limit $\delta\to 0$. Therefore we can take this limit and rewrite Eq.~(\ref{delta1-nonrel-LDOS})
as follows:
\begin{equation}
\label{nonrel-LDOS}
\begin{split}
\Delta N_\eta^{\mathrm{S}} & (\mathbf{r},E,B) = -\frac{M}{(\pi\hbar)^2}\frac{\sin\pi\eta}{2 \pi} \\
& \times \mbox{Im} \left[I \left(y,z \to - \frac{E+i \Gamma}{E_0},\eta \right) \right],
\end{split}
\end{equation}
where
\begin{equation}
\label{I-def}
\begin{split}
I & (y,z,\eta)=\int_0^\infty  d\beta e^{-\beta z}
\frac{e^{-y\coth\beta}}{\sinh\beta} \\
& \times \int_{-\infty}^\infty d\omega e^{-y\cosh\omega/\sinh\beta}
\frac{e^{-\eta(\omega+\beta)}}{1+e^{-(\omega+\beta)}},
\end{split}
\end{equation}
and the variable $y$ describes the spatial dependence.
Although the integrals in Eq.~(\ref{I-def}) can be evaluated numerically, this computation becomes
troublesome when the analytic continuation from $z>0$ to the complex values  $z \to - (E+i \Gamma)/E_0$
is done before the numerical integration. Thus our purpose is to derive such a representation
for $I(y,z,\eta)$ that it can be easily computed after the analytic continuation is done.
The function $I(y,z,\eta)$ is found in the Appendix~\ref{sec:D} and is given by
\begin{equation}
\label{I-final}
\begin{split}
& I(y,z,\eta)=\\
&\Gamma\left(\frac{z+1}{2}\right)\Gamma\left(\frac{z+2\eta-1}{2}\right)
             F_{(1-z-\eta)/2,(1-\eta)/2}(y)\\
+ & \Gamma\left(\frac{z-1}{2}\right)\Gamma\left(\frac{z+2\eta-1}{2}\right)
F_{(2-z-\eta)/2,\eta/2}(y),
\end{split}
\end{equation}
where the function $F_{\lambda,\mu}(y)$ is given by Eq.~(\ref{F-result}).

\begin{figure}[h]
\centering{
\includegraphics[width=0.5\textwidth]{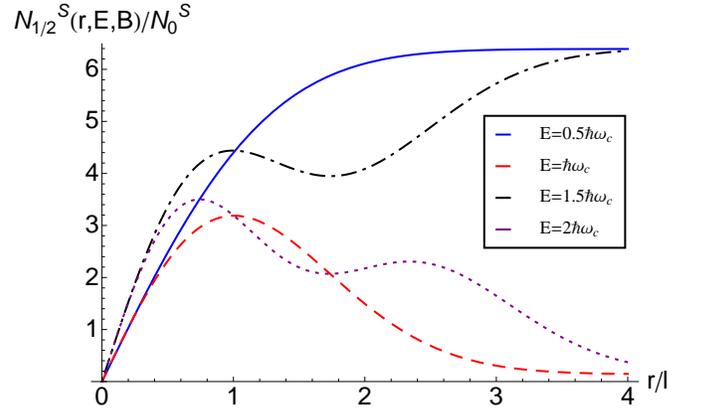}}
\caption{(Color online) The normalized full LDOS
$N_{1/2}^{\mathrm{S}}(r,E,B)/N_0^{\mathrm{S}}$ as a function of the distance $r$ measured in
the units of the magnetic length $l$ for four values of $E/\hbar \omega_c = 0.5, 1.5$ (usual LLs)
and $E/\hbar \omega_c = 1, 2$ (vortex-like levels).
The width is $\Gamma = 0.05 \hbar \omega_c$.}
\label{fig:2}
\end{figure}

The results of the numerical computation of the LDOS on the basis of Eqs.~(\ref{nonrel-LDOS})
and (\ref{I-final}) are shown in  Figs.~\ref{fig:1} and \ref{fig:2}. We emphasize that in Fig.~\ref{fig:1},
we plot the {\em full\/} LDOS $N_{\eta}^{\mathrm{S}}(r,E,B)$ as a function of energy $E$ for fixed values of $r$,
and in Fig.~\ref{fig:2}, the same quantity is presented as a function of the
distance $r$ from the vortex center for fixed values of $E$. Since Eq.~(\ref{nonrel-LDOS})
describes the perturbation of the LDOS $\Delta N_{\eta}^{\mathrm{S}}(r,E,B)$ by the vortex, to obtain the
value of the full LDOS $N_{\eta}^{\mathrm{S}}(r,E,B)$, we add to $\Delta N_{\eta}^{\mathrm{S}}$ its $\eta=0$
value, which is given by Eq.~(\ref{LDOS-nonrel-const}).
We note that in contrast to Ref.~\onlinecite{Slobodeniuk2010PRB}, when plotting these figures, we did not take into
account the presence of the finite carried density in 2DEG by shifting the energy origin. This makes more
straightforward a comparison with the Dirac case, where low carried densities are indeed accessible experimentally.

Although the model we consider is suitable for all values
of the distance from the center of the vortex $r$, there are obvious physical limitations on the possible
value of $r$ if the vortex penetrating graphene is coming from a type-II superconductor.
First of all, $r$ cannot be smaller than the vortex core, which is at least on the order of magnitude larger than
the distance scale $r_0$ of the order of the lattice constant.
We remind that in the previous paper,\cite{Slobodeniuk2010PRB} the distance $r$
was measured in the units of $r_0$, because for $B=0$ there is no such  natural scale as a magnetic length.
Secondly, we replace the magnetic field created by the other vortices by a constant background
magnetic field. This approximation may be appropriate if one considers a vicinity of the selected vortex,
which implies that $r$ has to be less than the intervortex distance $l_v$. This distance is
proportional to  the magnetic length,\cite{Tinkham.book} $l_v = c \sqrt{\pi} l \approx 1.77 l$, where
$c \approx 1$ is the geometric factor dependent on the Abrikosov's lattice structure. Thus although one can
investigate the regime $r \gg l$ theoretically, in practice it is not accessible.

In Fig.~\ref{fig:1}~(a)  we compare the already discussed after Eq.~(\ref{LDOS-nonrel-const})
case of the constant magnetic field  with the case when the Abrikosov vortex
is also present ($\eta=1/2$) for $r=l$. We observe that while for $\eta=0$
[the dashed (red) curve is, obviously,  $r$-independent] only the peaks at half-integers
$E/\hbar \omega_c$ are present, for $\eta=1/2$ the weight of these peaks is reduced and a set
of the new peaks at the integers $E/\hbar \omega_c$ on the solid (blue) curve is developed.
This behavior can be foreseen from the  expression for the full DOS difference (\ref{DOS-nonrel-vortex-const-delta})
[or Eq.~(\ref{DOS-nonrel-vortex-const})] discussed in Sec.~\ref{sec:LDOS-nonrel}.
The case with the Abrikosov vortex is further explored in Fig.~\ref{fig:1}~b, where
we plot the energy dependence of the LDOS  for $r = 0.5 l$ [the solid (blue) curve] and $r = 5 l$
[the dashed (red) curve]. Comparing the results for $r/l =0.5, 1.$, and $5.0$ we find that as the distance $r$
decreases, the integer  $E/\hbar \omega_c$ peaks are getting stronger, while for $r = 5.0 l$ they practically
disappear. This behavior allows to attribute the corresponding energy levels to the vortex.
On the other hand, the half-integer $E/\hbar \omega_c$ peaks corresponding to the usual LLs
(\ref{Schrodinger-LL}) formed in a constant magnetic field are getting weaker as the distance $r$
decreases. We stress that even for an arbitrary
vortex flux $\eta$, the latter levels will not change the positions, while the levels related to the vortex
will shift their energies.

Analyzing Eq.~(\ref{I-final}), which was used to plot Fig.~\ref{fig:1}, we observe that the positions of all
peaks are controlled by the gamma functions $\Gamma(z)$ which contain simple poles for $z=0,-1,-2,\ldots$.
However, the intensity of the peaks depends on the rather complicated modulating function $F_{\lambda,\mu}(y)$.
For example, we verified that despite that the gamma function $\Gamma[(z-1)/2]$ in the second term of Eq.~(\ref{I-final})
contains the pole at the negative energy $E = - \hbar \omega_c/2$, the final LDOS does not contain this pole.
To gain more insight on the behavior of the LDOS we have investigated its behavior in the limits $r \to 0$
and $r \to \infty$. Taking into account the $y \to 0$ limit of $\mbox{Im} I$ given by Eq.~(\ref{I-small-y}),
we obtain that the value $\Delta N_\eta^{\mathrm{S}}  (\mathbf{r}=0,E,B)$ is equal to the negative LDOS
(\ref{LDOS-nonrel-const}) in the constant magnetic field. This implies that the full LDOS in the center of the vortex
is completely depleted,
\begin{equation}
\label{LDOS-r=0}
N_\eta^{\mathrm{S}}  (\mathbf{r}=0,E,B)=0.
\end{equation}
Formally, this property reflects a simple fact that all solutions (\ref{solution-nonrel}) of the Schr\"{o}dinger
equation are vanishing at the origin, $\psi_{n,m}(r=0,\varphi)=0$.
This vortex-induced depletion of the LDOS in the nonrelativistic 2DEG was already seen
in  Ref.~\onlinecite{Slobodeniuk2010PRB} and now we conclude that it should also occur
in the presence of the background magnetic field.
This is exactly what we observe in Fig.~\ref{fig:2}, where all four curves begin from zero.
Two of these curves, viz. the solid (blue) and the dash-dotted (black) are for the usual LLs
with $E/\hbar \omega_c = 0.5, 1.5$, and  the other two [dashed (red) and dotted (violet)] are for the vortex
levels with $E/\hbar \omega_c = 1,2$.
For small $r < l$ all curves increase linearly as expected from the analytic results described in Appendix~\ref{sec:C}
if we take there $\eta=1/2$. Since for the large $y$ the function $F_{\lambda \mu}$ decays exponentially
[see Eq.~(\ref{F-large-y})], the LDOS difference  $\Delta N_\eta^{\mathrm{S}}  (\mathbf{r},E,B) \sim e^{-r^2/2l^2}$
for $r \to \infty$.
Accordingly, the large $r$ behavior of the full LDOS depends on the contribution of the position independent LDOS (\ref{LDOS-nonrel-const}).
Thus the large $r$ limit of all curves in Fig.~\ref{fig:2} is determined by the corresponding
value of the LDOS in the dashed (red) curve in Fig.~\ref{fig:1}~(a).

\section{Relativistic  case}
\label{sec:Dirac}
In Sec.~\ref{sec:Dirac-solutions}, we consider the solutions of the Dirac equation
\begin{equation}
\label{Dirac-eq}
H_D \Psi(\mathbf{r},\zeta) = E \Psi(\mathbf{r},\zeta),
\end{equation}
where the wave function is now a spinor
\begin{equation}
\Psi(\mathbf{r},\zeta) = \left[
                     \begin{array}{c}
                       \psi_1(\mathbf{r},\zeta) \\
                       \psi_2(\mathbf{r},\zeta) \\
                     \end{array}
                   \right],
\end{equation}
and the index $\zeta$ labels two inequivalent $\mathbf{K}_{\pm}$ points.
Notice that in Appendix~\ref{sec:E}, the definition (\ref{psi-def-appendix}) for
$\psi_2$ explicitly includes the factor $i$. Using these solutions in Sec.~\ref{sec:DOS-rel},
we obtain the full and the local DOSs that is considered in Sec.~\ref{sec:LDOS-rel}.

\subsection{Solutions of the Dirac equation and general representation for the  local density of states
and its limiting $\eta=0$ case}
\label{sec:Dirac-solutions}

The Dirac equation~(\ref{Dirac-eq}) with the regularized potential (\ref{reg-potential}) is solved in Appendix~\ref{sec:E}.
A general strategy is the same as described in Sec.~\ref{sec:nonrel-solutions}, but the main difference is in the matching
conditions.  While the radial components of the spinor $\Psi(r)$
have to be continuous:
\begin{equation}
\label{continuity}
\begin{split}
& \psi_1(R+0,\zeta)=\psi_1(R-0,\zeta), \\
& \psi_2(R+0,\zeta)=\psi_2(R-0,\zeta),
\end{split}
\end{equation}
their derivatives in contrast to the nonrelativistic case have a discontinuity:
\begin{equation}
\label{discontinuity}
\begin{split}
& \psi_1'(R+0,\zeta)-\psi_1'(R-0,\zeta)=\frac{\zeta\eta}{R}\psi_1(R,\zeta), \\
& \psi_2'(R+0,\zeta)-\psi_2'(R-0,\zeta)=-\frac{\zeta\eta}{R}\psi_2(R,\zeta).
\end{split}
\end{equation}
The discontinuity of the conditions (\ref{discontinuity}) follows from Eq.~(\ref{components-polar}) with the discontinuous
potential (\ref{reg-potential}). Another way to apprehend this discontinuity is to obtain a singular ar $r=R$
pseudo-Zeeman term squaring the Dirac equation (see Ref.~\onlinecite{Slobodeniuk2010PRB} for an overview).

After the limit $R\to 0$ is taken, we obtain the following solutions:
\begin{equation}
\label{Dirac:sol-m>0}
\begin{split}
& \Psi^{(\pm)}_{n,m}(\mathbf{r},1)= \\
& \frac{1}{2l\sqrt{\pi \mathcal{E}_{n,m}}}\left[
  \begin{array}{cc}
   \sqrt{\mathcal{E}_{n,m}\pm\Delta}\,e^{i(m-1)\varphi} J^n_{m+\eta-1}(y)\\
    \pm i \sqrt{\mathcal{E}_{n,m}\mp\Delta}\,e^{im\varphi} J^n_{m+\eta}(y)
  \end{array}
\right]
\end{split}
\end{equation}
for $m>0$,
\begin{equation}
\label{Dirac:sol-m=0}
\begin{split}
& \Psi^{(\pm)}_{n,0}(\mathbf{r},1)= \\
& \frac{1}{2l\sqrt{\pi \mathcal{E}_{n,0}}}\left[
  \begin{array}{cc}
   \sqrt{\mathcal{E}_{n,0}\pm\Delta}\,e^{-i\varphi} J^n_{1-\eta}(y)\\
    \mp i \sqrt{\mathcal{E}_{n,0}\mp\Delta}\,J^{n+1}_{-\eta}(y)
\end{array} \right]
\end{split}
\end{equation}
for $m=0$, and
\begin{equation}
\label{Dirac:sol-m<0}
\begin{split}
& \Psi^{(\pm)}_{n,m}(\mathbf{r},1)= \\
& \frac{1}{2l\sqrt{\pi \mathcal{E}_{n,m}}}\left[
  \begin{array}{cc}
\sqrt{\mathcal{E}_{n,m}\pm\Delta}\,e^{i(m-1)\varphi} J^n_{|m+\eta-1|}(y)\\
\mp i \sqrt{\mathcal{E}_{n,m}\mp\Delta}\,e^{im\varphi}J^{n+1}_{|m+\eta|}(y)
\end{array} \right]
\end{split}
\end{equation}
for $m<0$.
Here, the upper and lower signs $\pm$ correspond to the positive and negative energy solutions,
$\mathcal{E}^{(\pm)} =\pm \mathcal{E}_{n,m} $ with the absolute value of the energy
\begin{equation}
\begin{split}
\label{spectrum-rel}
& \mathcal{E}_{n,m}=\sqrt{\Delta^2+ \epsilon_0^2 \lambda_{n,m}}, \\
& \lambda_{n,m} = 2n+|m+\eta-1|+m+\eta+1
\end{split}
\end{equation}
for $n \geq 0$, the function $J_\nu^n(y)$ is defined by Eq.~(\ref{J-def}) with $y$ as in the nonrelativistic case,
and the relativistic Landau scale $\epsilon_0$ is defined after Eq.~(\ref{Dirac-LL}).
The zero-mode solution with $\mathcal{E} = - \Delta$ is a holelike
\begin{equation}
\label{Dirac:sol-n=0}
\Psi^{(-)}_{0,m}(\mathbf{r},1)=\frac{1}{\sqrt{2\pi}l}\left[
  \begin{array}{cc}
  0\\
  e^{im\varphi}\,J^0_{|m|-\eta}(y)
  \end{array}
\right], \quad m \leq 0.
\end{equation}
Let us now compare the solutions of the Schr\"{o}dinger and Dirac equations with the zero azimuthal number, $m=0$.
One can see from Eq.~(\ref{solution-nonrel}) that $\psi_{n,0}(r,1) \sim r^\eta$, because\cite{Bateman.book2}
\begin{equation}
L_n^\alpha(0) = \left(
                  \begin{array}{c}
                    n+ \alpha \\
                    n \\
                  \end{array}
                \right)
                = \frac{\Gamma(\alpha+n+1)}{\Gamma(\alpha+1) n!}.
\end{equation}
On the other hand, from Eqs.~(\ref{Dirac:sol-m=0}) and (\ref{Dirac:sol-n=0}) for $m=0$ we observe that
while the upper components are regular at $r=0$, the lower components diverge as  $\psi_{2 n,0}(r,1) \sim r^{-\eta}$.
Comparing these results with the behavior of the wave function in the Aharonov-Bohm field\cite{Slobodeniuk2010PRB}
we observe that the presence of the background magnetic field does not change the asymptotics of the $m=0$ solutions
for $r \to 0$. Also as expected,\cite{Slobodeniuk2010PRB} the zero-mode solution (\ref{Dirac:sol-n=0}) for
$\zeta=1$ and chosen direction of the field is holelike, $\mathcal{E}=-\Delta$.

The solutions for the case $\zeta=-1$ are the following:
\begin{equation}
\begin{split}
& \Psi^{(\pm)}_{n,m}(\mathbf{r},-1)=\\
& \frac{1}{2l\sqrt{\pi \mathcal{E}_{n,m}}}\left[
  \begin{array}{cc}
   \mp\sqrt{\mathcal{E}_{n,m}\pm\Delta}\,e^{im\varphi} J^n_{m+\eta}(y)\\
   i\sqrt{\mathcal{E}_{n,m}\mp\Delta}\,e^{i(m-1)\varphi} J^n_{m+\eta-1}(y)
  \end{array}
\right]
\end{split}
\end{equation}
for $m>0$,
\begin{equation}
\label{Dirac:sol-m=0-prime}
\begin{split}
& \Psi^{(\pm)}_{n,0}(\mathbf{r},-1)=\\
& \frac{1}{2l\sqrt{\pi \mathcal{E}_{n,0}}}\left[
  \begin{array}{cc}
   \pm\sqrt{\mathcal{E}_{n,0}\pm\Delta}\,J^{n+1}_{-\eta}(y)\\
    i \sqrt{\mathcal{E}_{n,0}\mp\Delta}\,e^{-i\varphi} J^n_{1-\eta}(y)
\end{array}
\right]
\end{split}
\end{equation}
for $m=0$, and
\begin{equation}
\begin{split}
& \Psi^{(\pm)}_{n,m}(\mathbf{r},-1)= \\
& \frac{1}{2l\sqrt{\pi \mathcal{E}_{n,m}}}\left[
  \begin{array}{cc}
\pm\sqrt{\mathcal{E}_{n,m}\pm\Delta}\,e^{im\varphi}J^{n+1}_{|m+\eta|}(y)\\
i \sqrt{\mathcal{E}_{n,m}\mp\Delta}\,e^{i(m-1)\varphi} J^n_{|m+\eta-1|}(y)
\end{array}
\right]
\end{split}
\end{equation}
for $m < 0$. Again the signs $\pm$ correspond to the solutions $\mathcal{E}^{(\pm)} = \pm \mathcal{E}_{n,m}$  with $n\geq 0$
and the energy $\mathcal{E}_{n,m}$ given by Eq.~(\ref{spectrum-rel}). Now the zero-mode  solution
\begin{equation}
\label{Dirac:sol-n=0-prime}
\Psi_{0,m}(\mathbf{r},-1)=\frac{1}{\sqrt{2\pi}l}\left[
  \begin{array}{cc}
  e^{im\varphi}\,J^0_{|m|-\eta}(y)\\
  0
  \end{array}
\right], \quad m \leq 0,
\end{equation}
is electron-like, $\mathcal{E} = \Delta$. Also the lower components of the $m=0$ solutions
(\ref{Dirac:sol-m=0-prime}) and (\ref{Dirac:sol-n=0-prime}) are regular at $r=0$,
and the upper components diverge as $\psi_{1 n,0}(r,-1) \sim r^{-\eta}$.

Since the solutions of the Dirac equation are characterized not only by the quantum numbers, but also by
the sublattice label $A$ and $B$, energy $\pm$, and the valley index $\zeta=\pm 1$,
instead of directly writing an analog of Eq.~(\ref{LDOS-def}),
it is more convenient to construct the Green's function expressing the LDOS
via the combinations of its matrix elements. The eigenfunction expansion for the retarded Green's function
reads
\begin{equation}
\begin{split}
\label{Dirac-GF}
G_\eta^{\mathrm{D}}(\mathbf{r},\mathbf{r}',E+i0;\zeta)= & \sum_{n=0}^{\infty} \sum_{m=-\infty}^\infty
\left(\frac{\Psi^{(+)}_{n,m}(\mathbf{r},\zeta)\Psi_{n,m}^{(+)\dag}(\mathbf{r}',\zeta) }{E-\mathcal{E}_{n,m}+i0} \right.\\
& \left. +\frac{\Psi^{(-)}_{n,m}(\mathbf{r},\zeta)\Psi_{n,m}^{(-)\dag}(\mathbf{r}')}{E+\mathcal{E}_{n,m}+i0}\right).
\end{split}
\end{equation}
The LDOS for $A$ and $B$ sublattices is expressed in terms of the Green's function (\ref{Dirac-GF})
as follows:
\begin{equation}
\label{LDOS-AB}
\begin{split}
N_\eta^{\mathrm{D} (A)}(\mathbf{r},E)=-\frac{1}{\pi} & \mbox{Im}
\left[   G_{\eta 11}(\mathbf{r},\mathbf{r},E+i\Gamma;\zeta=1)  \right. \\
+ & \left. G_{\eta11}(\mathbf{r},\mathbf{r},E+i\Gamma;\zeta=-1)   \right], \\
N_\eta^{\mathrm{D} (B)}(\mathbf{r},E)=-\frac{1}{\pi} & \mbox{Im}
\left[  G_{\eta 22}(\mathbf{r},\mathbf{r},E+i\Gamma;\zeta=1) \right.  \\
+ & \left. G_{\eta22}(\mathbf{r},\mathbf{r},E+i\Gamma;\zeta=-1)  \right],
\end{split}
\end{equation}
where similarly to the nonrelativistic case, the LL width $\Gamma$ is introduced.
Substituting the solutions of the Dirac equation in the Green's function
(\ref{Dirac-GF}), and using the definition (\ref{LDOS-AB}), we obtain
\begin{equation}
\label{LDOS-AB-1}
\begin{split}
N_\eta^{\mathrm{D} (A,B)} & (\mathbf{r},E,B) = - N_0^{\mathrm{D}}\frac{1}{\pi} \mbox{Im}
\left[ \frac{E + i \Gamma \pm \Delta}{\epsilon_0} \right. \\
& \times \left. G \left(y, z \to - \frac{(E + i \Gamma)^2 - \Delta^2}{\epsilon_0^2}, \eta\right) \right],
\end{split}
\end{equation}
where the upper (lower) sign corresponds to $A$ ($B$) sublattice, the relativistic Landau scale $\epsilon_0$ is
defined below Eq.~(\ref{Dirac-LL}),  and the normalization constant
$N_0^{\mathrm{D}} =\epsilon_0/(2 \pi\hbar^2 v_F^2)$ corresponds to the value of the free ($\eta=B=\Gamma=0$)
DOS for the Dirac quasiparticles per spin  and one sublattice (or valley)  taken at the energy
$E = \epsilon_0$ when the energy gap $\Delta=0$. Writing Eq.~(\ref{LDOS-AB-1}), we introduced the function
\begin{equation}
\label{G-via-gi}
G(y, z, \eta) = \sum_{i=1}^{3} g_i(y, z, \eta),
\end{equation}
which consists of the three terms
\begin{equation}
\label{g_i}
\begin{split}
& g_1(y, z, \eta) =-\sum_{n=0}^\infty\sum_{m=-\infty}^\infty
\frac{[J_{|m+\eta-1|}^n(y)]^2}{z+ \lambda_{n,m}}, \\
& g_2(y, z, \eta) = -\sum_{n=0}^\infty\sum_{m=-\infty}^\infty
\frac{[J_{|m+\eta|}^n(y)]^2}{z+\lambda^{\prime}_{n,m}},\\
& g_3(y, z, \eta) = \sum_{n=0}^\infty \left( \frac{[J_{\eta}^n(y)]^2}{z + 2(n+\eta)}
-\frac{[J_{-\eta}^n(y)]^2}{z+2n} \right)
\end{split}
\end{equation}
with  $\lambda_{n,m}$ defined in Eq.~(\ref{spectrum-rel}) and
\begin{equation}
\lambda^\prime_{n,m} = 2n+|m+\eta|+m+\eta.
\end{equation}
Note that the function $g_3$ contains the singular terms that originate from $m=0$
solutions of the Dirac equations.

The $g_{1,2}$ contributions are calculated in the same way as was derived Eq.~(\ref{delta1-nonrel-LDOS})
in Appendix~\ref{sec:A}, viz., exponentiating the denominators [see Eq.~(\ref{frac2exp})] and introducing
the regularizing factor $\delta >0$, and then using the sum (\ref{Lauerre-sum}), we obtain
\begin{equation}
\label{g1-Bessel}
\begin{split}
g_1(y,z,\eta)&=-\int_0^\infty d\beta e^{-\beta z}\frac{e^{-2(\delta+\beta)}}{1-e^{-2(\delta+\beta)}}e^{-y\coth(\delta+\beta)}\\
& \times \sum_{m=-\infty}^\infty
e^{-(\delta+\beta)(m+\eta)}I_{|m+\eta|}\left(\frac{y}{\sinh(\delta+\beta)}\right),
\end{split}
\end{equation}
where we also shifted the dummy index $m \to m+1$. As we saw in the nonrelativistic case, the presence of
$\delta$ is necessary for the calculation of the DOS, although it can be omitted in the expressions for the LDOS.
The remaining sum over $m$ can be found using
Eq.~(\ref{sum-m-final}) from Appendix~\ref{sec:B}
\begin{equation}
\label{Delta-g1}
\begin{split}
& \Delta g_1(y,z,\eta) \equiv g_1(y,z,\eta)-g_1(y,z,0)  \\
& =\frac{\sin\pi\eta}{\pi}\int_0^\infty
d\beta e^{-\beta z}\frac{e^{-2(\delta+\beta)}}{1-e^{-2(\delta+\beta)}} e^{-y\coth(\delta+\beta)} \\
& \times \int_{-\infty}^\infty d\omega e^{-y\cosh\omega/\sinh(\delta+\beta)}\frac{e^{-\eta(\omega+\delta+\beta)}}{1+e^{-(\omega+\delta+\beta)}},
\end{split}
\end{equation}
where we introduced the function $\Delta g_1$, which describes the perturbation by the vortex.
Similarly, for $\Delta g_2$ we have
\begin{equation}
\label{Delta-g2}
\begin{split}
& \Delta g_2(y,z,\eta) \equiv g_2(y,z,\eta)- g_2(y,z,0)\\
& =\frac{\sin\pi\eta}{\pi}\int_0^\infty
d\beta e^{-\beta z}\frac{1}{1-e^{-2(\delta+\beta)}} e^{-y\coth(\delta+\beta)} \\
& \times \int_{-\infty}^\infty d\omega e^{-y\cosh\omega/\sinh(\delta+\beta)}\frac{e^{-\eta(\omega+\delta+\beta)}}{1+e^{-(\omega+\delta+\beta)}}.
\end{split}
\end{equation}
The case of $g_3$ is even simpler, because there is no summation over $m$. Using the sum (\ref{Lauerre-sum})
we obtain an analog of Eq.~(\ref{g1-Bessel}). It contains the difference of two modified Bessel functions,
which can be expressed via the MacDonald function\cite{Bateman.book2}
\begin{equation}
K_\nu(x)=\frac{\pi}{2\sin\pi\nu}[I_{-\nu}(x)-I_\nu(x)].
\end{equation}
Finally we arrive at the result
\begin{equation}
\label{g3-MacDonald}
\begin{split}
& g_3(y,z,\eta)= -\frac{2\sin\pi\eta}{\pi} \\
& \times \int_0^\infty d\beta e^{-\beta z}
\frac{e^{-(\delta+\beta)\eta}}{1-e^{-2(\delta+\beta)}}e^{-y\coth(\delta+\beta)}K_\eta\left(\frac{y}{\sinh(\delta+\beta)}\right).
\end{split}
\end{equation}
Notice that since $g_3(y,z,\eta=0)=0$, there is no need to introduce a function $\Delta g_3$. Having the functions
$\Delta g_{1,2}$ and $g_3$ we can directly calculate the LDOS perturbation by the vortex,
$\Delta N_\eta^{\mathrm{D} (A,B)} (\mathbf{r},E,B) = N_\eta^{\mathrm{D} (A,B)} (\mathbf{r},E,B) -N_0^{\mathrm{D} (A,B)} (\mathbf{r},E,B)$.

\begin{figure}[h]
\centering{
\includegraphics[width=0.5\textwidth]{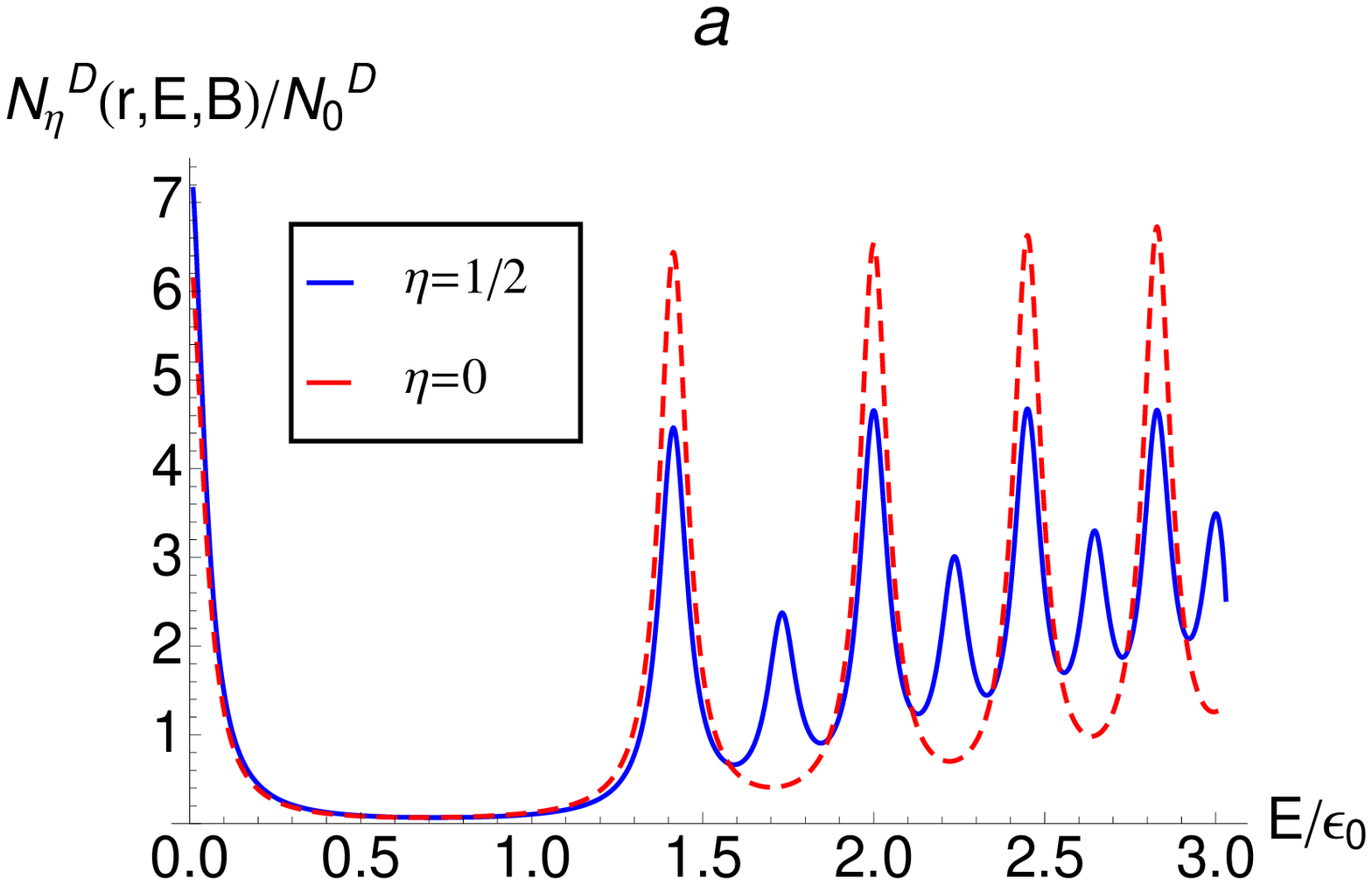}}
\centering{
\includegraphics[width=0.5\textwidth]{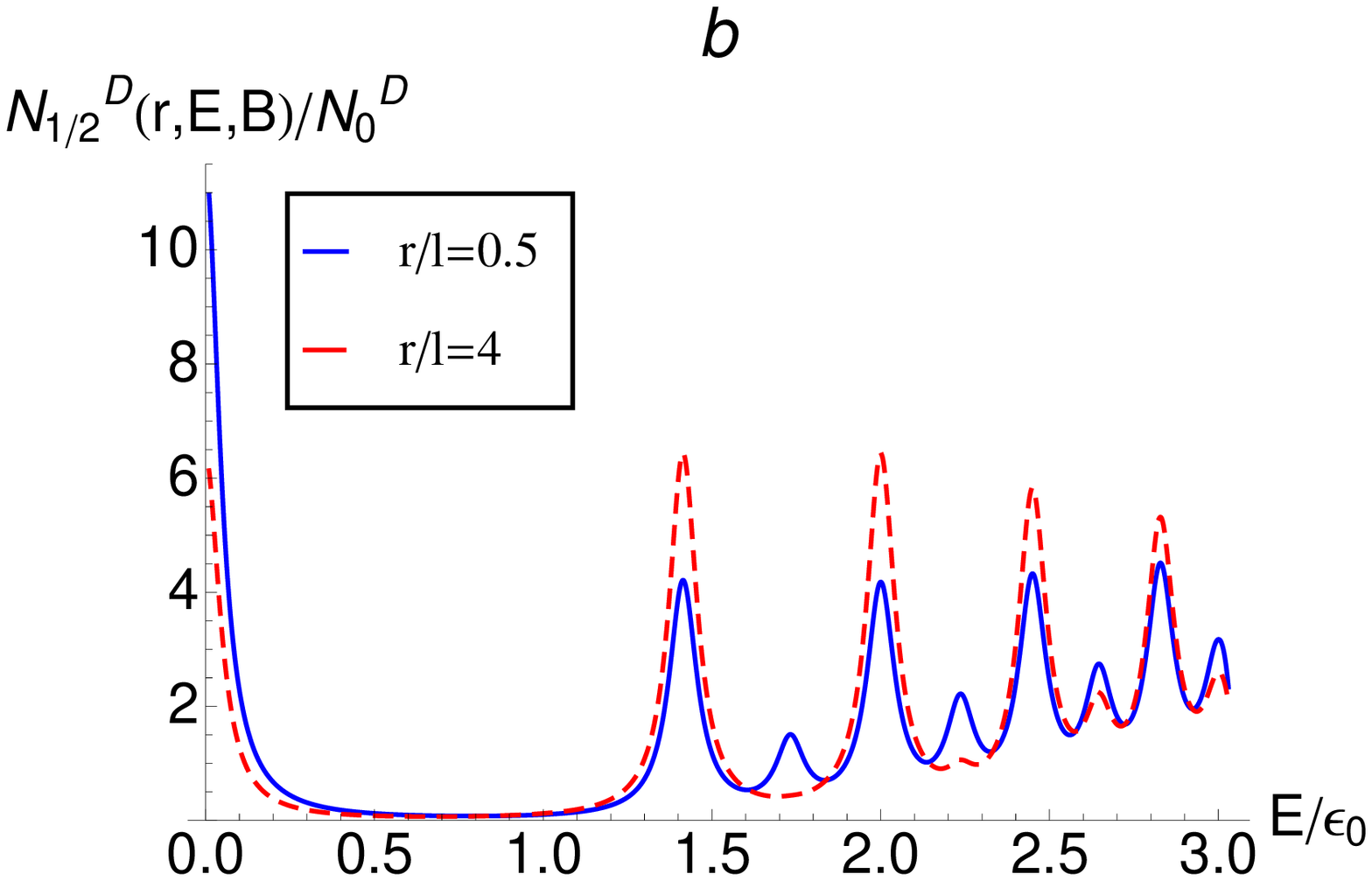}}
\caption{(Color online)
The normalized full LDOS $N_{\eta}^{\mathrm{D}}(r,E,B)/N_0^{\mathrm{D}}(\epsilon_0)$
as a function of energy $E$ in the units of the relativistic Landau scale $\epsilon_0$.
The LDOS is an even function of $E$, so only the positive-energy region is shown.
(a) $\eta = 0$ (no vortex and LDOS is $r$ independent) and $\eta=1/2$ for $r=l$. (b) Both lines are for $\eta=1/2$, $r=0.5 l$,
and $r = 4l$. In all cases, the width is $\Gamma=0.05\epsilon_0$, $W/\epsilon_0=3.35$, and $\Delta=0$.}
\label{fig:3}
\end{figure}

The subsequent consideration is made in parallel to the nonrelativistic case. We consider first
the LDOS in the constant magnetic field ($\eta=0$) when due to the translational invariance it coincides with
the DOS per unit area. The LDOS can be derived in a similar to Eq.~(\ref{LDOS-nonrel-const})
way, but a special care has to be taken because in contrast to the nonrelativistic case, the cutoff parameter
$\delta$ enters the final result,
\begin{equation}
\label{LDOS-rel-const-prefinal}
\begin{split}
& N_0^{\mathrm{D} (A,B)}   (E,B)= - \frac{N_0^{\mathrm{D}}}{\pi} \\
& \times \mbox{Im} \left[\frac{E+i\Gamma \pm \Delta}
{\varepsilon_0}
\left(\ln (2 \delta) + \gamma+ \psi\left(\frac{z}{2}\right)+\frac{1}{z}\right)\right],
\end{split}
\end{equation}
where we kept only the divergent in the limit $\delta \to 0$ terms
It is more convenient to  rewrite   Eq.~(\ref{LDOS-rel-const-prefinal}) in the form
derived in Ref.~\onlinecite{Sharapov2004PRB}, where instead of the cutoff $\delta$  the bandwidth $W$ cutoff is used
\begin{equation}
\label{LDOS-rel-const}
\begin{split}
& N_0^{\mathrm{D} (A,B)}  (E,B)=\frac{N_0^{\mathrm{D}}}{\pi}\left\{\frac{\Gamma}{\epsilon_0}
\ln\frac{W^2}{2 \epsilon_0^2}- {\rm Im}\left[ \frac{E+i\Gamma \pm \Delta}{\epsilon_0} \right. \right.
\\ & \left. \left. \times \left(\psi \left(\frac{\Delta^2-(E+i\Gamma)^2 }{2 \epsilon_0^2}\right)+
\frac{\epsilon_0^2}{\Delta^2-(E+i\Gamma)^2} \right)\right]\right\}.
\end{split}
\end{equation}
The advantage of the representation (\ref{LDOS-rel-const}) is that its $B\to 0$ limit takes the usual
form.\cite{Sharapov2004PRB}
The quantum magnetic oscillations of the LDOS,
$N_{0}^{\mathrm{D}}(E,B) = N_{0}^{\mathrm{D}(A)}(E,B) = N_{0}^{\mathrm{D}(B)}(E,B)$ for $\Delta=0$
are shown in Fig.~\ref{fig:3}~(a) on a dashed (red) curve. Only the positive-energy region is shown,
where the positions of the peaks, $\mathcal{E}_n/\epsilon_0 = \sqrt{2n}$, are in accord with the Dirac spectrum
(\ref{Dirac-LL}). The nonequidistant LLs along with the peak at $E=0$ related to the energy independent lowest LL
are characteristic of the Dirac fermions.
The reflection formula (\ref{psi-inv}) allows one to extract these oscillations analytically.\cite{Sharapov2004PRB}

The representation (\ref{LDOS-AB-1}) for the LDOS, where the function (\ref{G-via-gi}) consists of the three terms
(\ref{Delta-g1}),  (\ref{Delta-g2}), and (\ref{g3-MacDonald}) which describe the LDOS perturbation,
is our starting point for the analysis of the LDOS and DOS in the relativistic case.
In the next Sec.~\ref{sec:DOS-rel}, we begin with the DOS and in Sec.~\ref{sec:LDOS-rel}
return to the LDOS.

\subsection{The density of states}
\label{sec:DOS-rel}

The full DOS per spin projection is given by the spatial integral
(\ref{DOS-nonrel}). Accordingly, the full DOS perturbation by the vortex
$\Delta N_\eta^{\mathrm{D} (A,B)} (E,B) = N_\eta^{\mathrm{D} (A,B)} (E,B) -N_0^{\mathrm{D} (A,B)} (E,B)$
for $A$ and $B$ sublattices takes the form
\begin{equation}
\label{DOS-rel}
\begin{split}
& \Delta N_\eta^{\mathrm{D} (A,B)} (E,B) = - N_0^{\mathrm{D}}
2 l^2  \mbox{Im}
\left[\frac{E+i\Gamma \pm \Delta}{\varepsilon_0} \right. \times \\
& \left. \int_0^\infty dy (\Delta g_1(y,z,\eta) +
\Delta g_2(y,z,\eta) + g_3(y,z,\eta))\right].
\end{split}
\end{equation}
The subsequent calculation on the basis of Eq.~(\ref{DOS-rel}) is similar to the nonrelativistic
case considered in Appendix~\ref{sec:C}, and gives [compare Eqs.~(\ref{DOS-nonrel-vortex-const-z}) and (\ref{DOS-nonrel-vortex-const})]
\begin{equation}
\label{DOS-rel-vortex-const-psi}
\begin{split}
& \Delta N_\eta^{\mathrm{D} (A,B)} (E,B)=- \mbox{Im} \left\{ \frac{E+i\Gamma\pm\Delta}{2\pi\varepsilon_0^2}
\left[ 2\eta \left(1+\frac{1}{z} \right) \right. \right.\\
& \left. \left. + (z+2\eta) \left( \psi \left(\frac{z}{2}\right)-
\psi\left(\frac{z+2\eta}{2}\right) \right) \right] \right\},
\end{split}
\end{equation}
where $z \to - [(E + i \Gamma)^2 - \Delta^2]/\epsilon_0^2$.
In the clean limit $\Gamma \to 0$ the DOS difference reduces to
\begin{equation}
\label{DOS-rel-vortex-const-delta}
\begin{split}
\Delta & N_\eta^{\mathrm{D} (A,B)}  (E,B) = \eta \delta(E \pm \Delta) \\
& + 2(E \pm \Delta) \mbox{sgn} E \left[ \sum_{n=1}^\infty n \delta (E^2 - \Delta^2 - 2 (n + \eta) \epsilon_0^2 ) \right.\\
& \left. -\sum_{n=1}^{\infty} (n - \eta)\delta (E^2 - \Delta^2 - 2 n \epsilon_0^2 ) \right],
\end{split}
\end{equation}
where except of the first, proportional to $\eta$, zero-mode term
each $\delta$ function corresponds to both positive and negative energy peaks.
A comparison of this result with Eq.~(\ref{DOS-nonrel-vortex-const-delta}) for the nonrelativistic problem
sheds the light on the difference between these cases. We observed from Eq.~(\ref{DOS-nonrel-vortex-const-delta})
that all peaks associated with the usual LLs are depleted, while the peaks related to the vortex are developed.
At first sight, Eq.~(\ref{DOS-rel-vortex-const-delta}) follows the same pattern, viz. the LL
peaks with $\mathcal{E}_n^{(\pm)} = \pm \sqrt{\Delta^2 + 2 \epsilon_0^2 n }$ with $n=1,2,\dots$ are depleted and the vortex-like levels
$\mathcal{E}_n^{(\pm)} = \pm \sqrt{\Delta^2 + 2 \epsilon_0^2( n + \eta)}$ with $n=1,2,3,\dots$ are developed.
However, the  first term $\eta \delta(E \pm \Delta)$ related to the zero-mode solutions of the Dirac equation is present for
any magnetic field configuration and the addition of the vortex only adds $\eta$ to the weight of the corresponding peak.
This property is an illustration of the topological origin of the lowest LL.\cite{Jackiw1984PRD}

The $B \to 0$ limit can again be obtained using the asymptotic expansion
(\ref{psi-expand}), which for the expression in the square brackets of Eq.~(\ref{DOS-rel-vortex-const-psi})
gives
\begin{equation}
\label{DOS-expand}
\begin{split}
& 2\eta \left(1+\frac{1}{z}\right) + (z+2\eta) \left( \psi \left(\frac{z}{2}\right)-
\psi\left(\frac{z+2\eta}{2}\right) \right) \\
& = -\frac{2\eta^2}{z} + O\left(\frac{1}{z^2}\right).
\end{split}
\end{equation}
Substituting Eq.~(\ref{DOS-expand}) in  Eq.~(\ref{DOS-rel-vortex-const-psi}) and making the analytic
continuation $z \to - [(E + i \Gamma)^2 - \Delta^2]/\epsilon_0^2$
in the clean limit $\Gamma \to 0$, we obtain
\begin{equation}
\label{DOS-rel-depletion}
\begin{split}
\Delta N_\eta^{\mathrm{D}  (A,B)}(E,B=0) &= N_\eta^{\mathrm{D}  (A,B)}(E,B=0)- V_{2D} N_0^{\mathrm{D}  (A,B)} \\
& = \eta^2 \delta(E\mp \Delta).
\end{split}
\end{equation}
This result is in agreement with Refs.~\onlinecite{Moroz1995PLB} and \onlinecite{Slobodeniuk2010PRB},
where the DOS $\rho_\eta^{\mathrm{D}}(E,\zeta)$ for a separate
$\mathbf{K}_{\pm}$ point, but summed contributions for $A$ and $B$ sublattices, was considered. Its perturbation
$\Delta \rho_\eta^{\mathrm{D}}(E,\zeta) = \rho_\eta^{\mathrm{D}}(E,\zeta) - V_{2D} \rho_0^{\mathrm{D}}(E) $ with
respect to the free DOS per spin and one valley,  $\rho_0^{\mathrm{D}}(E) = |E| \theta (E^2/\Delta^2 - 1)/(2 \pi \hbar^2 v_F^2) $,
is equal to
\begin{equation}
\label{Dirac-DOS-final}
\begin{split}
\Delta \rho_\eta^{\mathrm{D}} (E,\zeta) = &-\frac{1}{2}\eta(1-\eta)[\delta(E-\Delta)+\delta(E+\Delta)] \\
&+ \eta \delta (E + \zeta  \Delta), \quad \eta >0.
\end{split}
\end{equation}

Integrating Eq.~(\ref{DOS-rel-depletion}) one can find the total excess of the states induced by the vortex
\begin{equation}
\label{excess}
\begin{split}
& \Delta N_\eta^{\mathrm{D}} \equiv \\
& \int_{-\infty}^{\infty}
d E (\Delta N_\eta^{\mathrm{D}(A)}(E,B)+ \Delta N_\eta^{\mathrm{D}(B)}(E,B))  = 2 \eta^2.
\end{split}
\end{equation}
As in the nonrelativistic case (\ref{deficit}), it turns out that the integral (\ref{excess})
does not depend on the strength $B$ of the background field. This
can be checked by integrating the sum  (\ref{DOS-rel-vortex-const-delta}) and using an appropriate regularization.
Completing our discussion of the DOS we note that the value $\Delta N_\eta^{\mathrm{D}}$ has to be
distinguished from the induced by the magnetic flux fractional fermion number\cite{Niemi1983PRL}
which in terms of the DOS (\ref{Dirac-DOS-final}) can be written as follows:
\begin{equation}
N_\eta = -\frac{1}{2}\int_{-\infty}^{\infty} d E \mbox{sgn} E \Delta \rho_\eta^{\mathrm{D}} (E,\zeta) = \frac{\zeta\eta}{2}.
\end{equation}

\subsection{The local density of states}
\label{sec:LDOS-rel}

The contributions $\Delta g_{1,2}$ to the relativistic LDOS given by Eqs.~(\ref{Delta-g1}) and (\ref{Delta-g2})
can be written in terms of the function $I(y,z,\eta)$
defined by Eq.~(\ref{I-def}), which was used in Sec.~\ref{sec:LDOS-nonrel} to express the nonrelativistic LDOS
\begin{equation}
\label{g_1via-I}
\Delta g_1(y,z,\eta)= \frac{\sin\pi\eta}{2\pi}I(y,z+1,\eta),
\end{equation}
and
\begin{equation}
\label{g_2via-I}
\Delta g_2(y,z,\eta)= \frac{\sin\pi\eta}{2\pi}I(y,z-1,\eta).
\end{equation}
Thus the only remaining term  we have to find is $g_{3}$ given by Eq.~(\ref{g3-MacDonald}).
Changing the variable $x=e^{-\beta}$, we obtain
\begin{equation}
\begin{split}
& g_3(y,z,\eta)= -\frac{2\sin\pi\eta}{\pi} \\
& \times \int_0^1 dx
\frac{x^{z+\eta-1}}{1-x^2}e^{-y(1+x^2)/(1-x^2)}K_\eta\left(\frac{2xy}{1-x^2}\right),
\end{split}
\end{equation}
Now using the integral (2.16.10.5) from\cite{Prudnikov.vol2} (one can also change the variable to $t$ via
$e^{-\beta}=[t/(1+t)]^{1/2}$ and use the integral (\ref{Prud-int})), we can write
\begin{equation}
\begin{split}
& \int_0^y dx \frac{x^{\alpha-1}}{x^2-y^2}\exp\left(-b\frac{y^2+x^2}{y^2-x^2}\right)
K_\nu\left(\frac{2cx}{y^2-x^2}\right)\\
 =&-\frac{y^{\alpha-1}}{4c}\Gamma\left(\frac{\alpha-\nu}{2}\right)
\Gamma\left(\frac{\alpha+\nu}{2}\right)\\
& \times W_{(1-\alpha)/2,\nu/2}\left(b+\sqrt{b^2-(c/y)^2}\right)\\
& \times W_{(1-\alpha)/2,\nu/2}\left(b-\sqrt{b^2-(c/y)^2}\right).
\end{split}
\end{equation}
Thus we can express $g_3$ in terms of the Whittaker function $W_{\lambda \mu}(z)$
as follows:
\begin{equation}
\label{g3-final}
g_3(y,z,\eta)=\frac{\sin\pi\eta}{2\pi} I^{\mathrm{D}}(y,z,\eta)
\end{equation}
with the function
\begin{equation}
\label{ID-def}
I^{\mathrm{D}}(y,z,\eta) = -\frac{1}{y}
\Gamma \left(\frac{z}{2} \right)\Gamma \left( \frac{z+2\eta}{2} \right)W_{(1-z-\eta)/2,\eta/2}^2(y).
\end{equation}
Thus the final expression for the relativistic LDOS perturbation by the vortex takes the form
\begin{equation}
\label{Delta-LDOS-AB}
\begin{split}
\Delta & N_\eta^{\mathrm{D} (A,B)}  (\mathbf{r},E,B) = - N_0^{\mathrm{D}}\frac{1}{\pi} \mbox{Im}
\left[ \frac{E + i \Gamma \pm \Delta}{\epsilon_0} \right. \\
& \times \left. \Delta G \left(y, z \to - \frac{(E + i \Gamma)^2 - \Delta^2}{\epsilon_0^2}, \eta\right) \right],
\end{split}
\end{equation}
where the function
\begin{equation}
\label{DGviaI}
\begin{split}
&\Delta G(y,z,\eta) = \frac{\sin \pi \eta}{2 \pi}\\
& \times [I(y,z+1,\eta)+I(y,z-1,\eta) + I^{\mathrm{D}}(y,z,\eta)]
\end{split}
\end{equation}
is expressed via the defined above functions (\ref{I-final}) and (\ref{ID-def}).

To complete the analytic treatment, we consider the behavior of the LDOS in the most interesting case
of the small $r$, when we expect that the difference between the relativistic and nonrelativistic cases
should be the most transparent. The observation (\ref{LDOS-r=0}) that in the nonrelativistic case the full LDOS
in the center of the vortex vanishes turns out to be useful for better understanding of the relativistic case.
Indeed, let's consider the first two terms of Eq.~(\ref{G-via-gi}) with $g_{1,2}$ that contribute to the full
LDOS (\ref{LDOS-AB-1}). Since the numerators of $g_{1,2}$ in Eq.~(\ref{g_i}) vanish at $y=0$,
the only term that governs the behavior of the full LDOS in the $r \to 0$ limit  is the function
$I^{\mathrm{D}}(y,z,\eta)$, which due to its origin from the $m=0$ solutions is expected to be divergent.

The same result can be verified using the final expressions (\ref{Delta-LDOS-AB}) and (\ref{DGviaI}) for
$\Delta N_\eta^{\mathrm{D} (A,B)}(\mathbf{r}=0,E,B)$. For $y=0$, the first two terms of
Eq.~(\ref{DGviaI}) with the function $I$, which originate from $\Delta g_{1,2}$ [see Eqs.~(\ref{g_1via-I}) and (\ref{g_2via-I})]
can be combined together;
\begin{equation}
\begin{split}
& \frac{\sin \pi \eta}{2 \pi} \mbox{Im} [I(y=0,z+1,\eta)+I(y=0,z-1,\eta) ] \\
& = - \mbox{Im}  \left[  \psi \left(\frac{z}{2}\right) + \frac{1}{z}\right],
\end{split}
\end{equation}
where we used the value $\mbox{Im} I(y = 0,z,\eta)$ established in Eq.~(\ref{I-small-y})
and then transformed the first digamma function using Eq.~(\ref{psi-recurrent}).
Thus we find that in the limit $\Gamma \to 0$ the contribution of these $\Delta g_{1,2}$
terms to the LDOS difference $\Delta N_\eta^{\mathrm{D}(A,B)}(\mathbf{r}=0,E,B)$ given by Eq.~(\ref{Delta-LDOS-AB})
is equal to the negative LDOS (\ref{LDOS-rel-const-prefinal})
in the constant magnetic field.

Let us now analyze the behavior of the function $I^{\mathrm{D}}(y,z,\eta)$
in the $r \to 0$ limit.
Using the expansion of the Whittaker  function
(\ref{Whittaker-small}) in the limit $y \to 0$, we obtain
\begin{equation}
\begin{split}
I^{\mathrm{D}}&(y,z,\eta) \\
&= - \frac{\Gamma(z/2) \Gamma^2(\eta)}{\Gamma(\eta+z/2)} y^{-\eta} + O(y^0), \quad y \to 0.
\end{split}
\end{equation}
Thus the full LDOS is divergent at the origin as
\begin{equation}
\label{LDOS-Dirac-small}
\begin{split}
& N_\eta^{\mathrm{D} (A,B)}  (\mathbf{r},E,B) \sim r^{-2\eta} \mbox{Im}
\left[ \frac{E + i \Gamma \pm \Delta}{\epsilon_0} \times \right.\\
& \left. \Gamma \left( \frac{ \Delta^2 - (E + i \Gamma)^2 }{2 \epsilon_0^2}\right)
\Gamma^{-1} \left( \frac{\Delta^2 - (E + i \Gamma)^2}{2 \epsilon_0^2} + \eta \right)\right].
\end{split}
\end{equation}
For $\eta=1/2$, the divergence is $\sim r^{-1}$ as was in the absence of the background field.\cite{Slobodeniuk2010PRB}
As we discuss below, the presence of this field makes the divergence of the LDOS strongly energy dependent.

\begin{figure}[h]
\centering{
\includegraphics[width=0.5\textwidth]{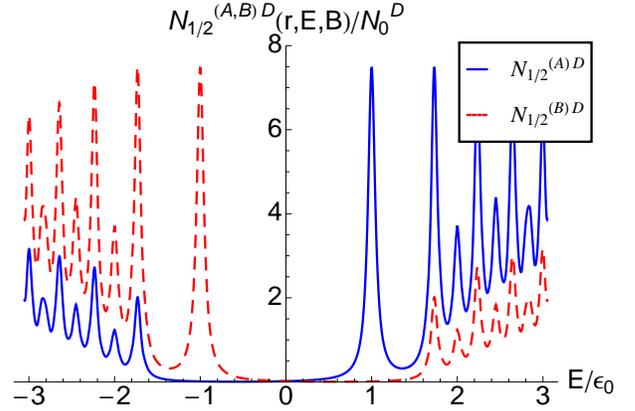}}
\caption{(Color online)
The normalized full LDOS $N_{1/2}^{\mathrm{D}(A,B)}(r,E,B)/N_0^{\mathrm{D}}(\epsilon_0)$
as a function of energy $E$ in the units of the relativistic Landau scale $\epsilon_0$ for $r=l$.
The gap is $\Delta=\epsilon_0$, the width is $\Gamma=0.05\epsilon_0$, and $W/\epsilon_0=3.35$.}
\label{fig:4}
\end{figure}

\begin{figure}[h]
\centering{
\includegraphics[width=0.5\textwidth]{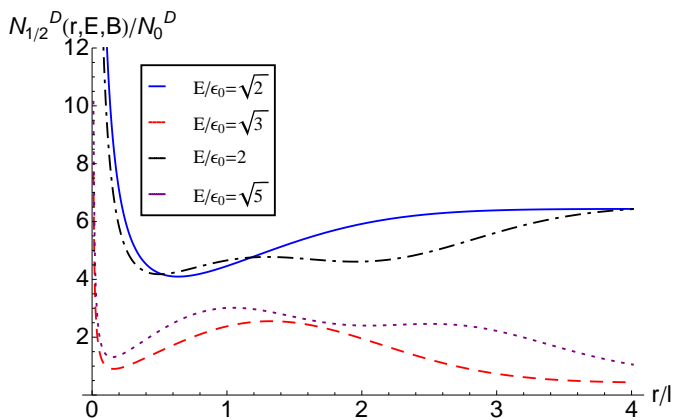}}
\caption{(Color online)
The  normalized full LDOS $N_{1/2}^{\mathrm{D}}(r,E,B)/N_0^{\mathrm{D}}(\epsilon_0)$
as a function of distance $r$ measured in the units of the magnetic length $l$
from the vortex  for four values of  $E/\epsilon_0= \sqrt{2},2$ (usual LLs) and
$E/\epsilon_0= \sqrt{3}, \sqrt{5}$ (vortex-like levels).
The width is $\Gamma=0.05\epsilon_0$, $W/\epsilon_0=3.35$, and $\Delta=0$.}
\label{fig:5}
\end{figure}

The results of the numerical computations of the {\em full\/} LDOS on the basis of Eqs.~(\ref{Delta-LDOS-AB})
and (\ref{DGviaI}) are shown in  Figs.~\ref{fig:3}, \ref{fig:4}, and \ref{fig:5}.
Since Eq.~(\ref{Delta-LDOS-AB}) describes the perturbation of the LDOS $\Delta N_{\eta}^{\mathrm{D}}(r,E,B)$ by
the vortex, to obtain the value of the full LDOS $N_{\eta}^{\mathrm{D}}(r,E,B)$, we add to $\Delta N_{\eta}^{\mathrm{S}}$
its $\eta=0$ given by Eq.~(\ref{LDOS-rel-const}).

In Fig.~\ref{fig:3}~(a) we compare the already discussed after Eq.~(\ref{LDOS-rel-const}) case of the LDOS for
a constant magnetic  field with the case when the vortex is also present ($\eta=1/2$) for $r=l$.
Since we consider the situation when $\Delta=0$, there is no difference between sublattices,
$N_{0}^{\mathrm{D}}(E,B) = N_{0}^{\mathrm{D}(A)}(E,B) = N_{0}^{\mathrm{D}(B)}(E,B)$ and the LDOS is an
even function of energy, so the positive energy region is plotted.
We observe that compared to $\eta=0$ [the dashed (red) curve] for $\eta=1/2$ [solid (blue) curve] a set of the new peaks
at $\mathcal{E}_n/\epsilon_0 = \sqrt{2n+1}$ with $n=1,2,\ldots$ is developed and the lowest LL peak ($n=0$) is enhanced.
This behavior can be foreseen from the expression for the full DOS difference Eq.~(\ref{DOS-rel-vortex-const-delta})
[or Eq.~(\ref{DOS-rel-vortex-const-psi})] discussed in Sec.~\ref{sec:DOS-rel}.
The case with the Abrikosov vortex is further explored in Fig.~\ref{fig:3}~(b), where
we plot the energy dependence of the LDOS  for $r = 0.5 l$ [the solid (blue) curve] and $r = 4 l$
[the dashed (red) curve]. Comparing the results for $r/l =0.5, 1.0$, and $4.0$ we find that as the distance $r$
decreases, the peaks at $\mathcal{E}_n/\epsilon_0 = \sqrt{2n+1}$ with $n=1,2,3\ldots$ related to the vortex
are getting stronger. When $r$ further decreases, the peaks related to LLs grow faster than the vortexlike peaks.
This behavior indeed allows to attribute the corresponding energy levels to the vortex.
On the other hand, the peaks
$\mathcal{E}_n/\epsilon_0 = \sqrt{2n}$ with $n=1,2,\ldots$ corresponding to the usual LLs
(\ref{Dirac-LL}) are getting weaker as the distance $r$ decreases.
We remind that even for an arbitrary
vortex flux $\eta$ the latter levels will not change the positions, while the levels related to the vortex
will shift their energies.
Fig.~\ref{fig:3} also illustrates a special character of the lowest LL that is present even in
an inhomogeneous magnetic field (see also recent simulations in Ref.~\onlinecite{Roy2011}),
and therefore is getting stronger as $r$ decreases.

In Fig.~\ref{fig:4} we consider the energy dependence of the LDOS $N_{1/2}^{\mathrm{D}(A,B)}(r,E,B)$
when there is a gap $\Delta = \epsilon_0$ in the spectrum.  The distance from the vortex center is $r=l$.
The gap introduces asymmetry between the
LDOS on $A$ and $B$ sublattices and also makes the LDOS asymmetric with respect to $E=0$, so we have to
plot both negative- and positive-energy regions. Indeed we observe that the zero LL peak at
$\mathcal{E} = \Delta$ is present only in $N_{1/2}^{\mathrm{D}(A)}(r,E,B)$, while the peak at
$\mathcal{E} = -\Delta$ shows up  only in  $N_{1/2}^{\mathrm{D}(B)}(r,E,B)$.
The vortexlike levels also become asymmetric with respect to  $E=0$.
All this illustrates that the STS on graphene on a substrate that can induce inequivalence of sublattices in
graphene should reveal these features.

From Eq.~(\ref{LDOS-Dirac-small}) we expect that the presence of the background
magnetic field makes $r^{-1}$ divergence at $r \to 0$ of the LDOS strongly energy dependent:
it is emphasized by the poles of the first $\Gamma$ function, when the energy $E$
is close to the energies of  the usual LLs,
$\mathcal{E}_n^{(\pm)} = \pm \sqrt{\Delta^2 + 2 \epsilon_0^2 n }$ with $n=0,1,2,\dots$, and
oppositely, because the second $\Gamma$ function is in the denominator, when $E$
is equal to the energies of the vortexlike levels,
$\mathcal{E}_n^{(\pm)} = \pm \sqrt{\Delta^2 + 2 \epsilon_0^2( n + \eta)}$ with $n=1,2,\dots$, the
divergence is suppressed.
This is exactly what we observe in Fig.~\ref{fig:5}, where we show the dependence of the LDOS on the
distance $r$ for fixed values of the energy ($\Delta=0$). Indeed,  the solid (blue)
and dash-dotted (black) curves which correspond to  energies $E/\epsilon_0 = \sqrt{2}, 2$ of the usual LLs
have divergent behavior at the origin.
Obviously, this divergence is also present for $E=0$ and the corresponding curve
will be above the higher energy curves $E/\epsilon_0 = \sqrt{2}$ and $E/\epsilon_0 = 2$.
On the other hand, the dashed (red) and  dotted (violet)  curves, which correspond to the energies
$E/\epsilon_0 = \sqrt{3}$, and $\sqrt{5}$ of the vortexlike levels tend to go to a constant
value at $r=0$. Strictly speaking $r^{-1}$ divergence is suppressed only when the function $\Gamma^{-1}(\eta+z/2)$
in Eq.~(\ref{LDOS-Dirac-small}) is zero, but for small values of the level width $\Gamma$, the
divergence seems to be completely suppressed for the chosen values of the energy.
For $r \gg l$, the behavior of the LDOS resembles the nonrelativistic case. Since in this limit
the LDOS difference $\Delta N_\eta^{\mathrm{S}}  (\mathbf{r},E,B) \sim e^{-r^2/2l^2}$, the large-$r$
behavior of the full DOS is determined by the contribution of the position independent LDOS
(\ref{LDOS-rel-const}). Thus the large $r$-limit of all curves in Fig.~\ref{fig:5}
is determined by the corresponding value of the LDOS in the dashed (red) curve in Fig.~\ref{fig:3}~(a).

\section{Conclusions}
\label{sec:concl}

The main motivation of this work was to address the question as to whether one can distinguish graphene from
2DEG by measuring the LDOS near the Abrikosov vortex penetrating them. In the first
publication,\cite{Slobodeniuk2010PRB} we investigated the simplest formulation of the problem with a single
vortex.
In the 2DEG, the solutions of the Schr\"{o}dinger equation in the presence of the Aharonov-Bohm
field are regular  and the LDOS near the vortex is depleted.
On the other hand, a specific feature of the  Dirac fermions in the field of the Aharonov-Bohm flux, namely,
such as the presence of the divergent as $r^{-\eta}$ at the origin  the $m=0$ solution of the Dirac equation,
results in the $r^{-2 \eta}$ divergence of the LDOS in the vicinity of the vortex. Therefore the LDOS enhancement
near the vortex can really distinguish graphene from 2DEG.

This positive answer obtained in the previous paper\cite{Slobodeniuk2010PRB}
is now extended for the case of a more complicated magnetic field configuration consisting of the Aharonov-Bohm flux
and a constant background field, as one can see just from a comparison of  Figs.~\ref{fig:2} and \ref{fig:5}.
It turns out that the character of the divergence in the Dirac case remains the same,
but it is now strongly modulated by the energy-dependent factor. The divergence is present when the energy
is equal to the energies of the usual LLs (\ref{Dirac-LL}), including the lowest zero-energy  LL.

The significant difference between the relativistic and nonrelativistic cases
can be understood by comparing the squared Dirac equation with the Schr\"{o}dinger equation.
While the Schr\"{o}dinger equation contains only an effective centrifugal potential,
which originates from the angular part of the Hamiltonian, an equation for one of the components
of the Dirac spinor always contains an attractive pseudo-Zeeman term. We call it
the pseudo-Zeeman term because it is related to the sublattice rather than to the spin degree of freedom.
Since for the zero azimuthal number $m$ the centrifugal part of the potential is the smallest,
the attraction term results in the divergence of the LDOS near the vortex.
Our main results, which allow to conclude that this picture remains valid in the constant background field
can be summarized as follows.

(i) We obtained analytic expression for the LDOS perturbation by the Aharonov-Bohm flux in the
presence of a constant background magnetic field in the nonrelativistic, see Eqs.~(\ref{nonrel-LDOS}) and (\ref{I-final}), and
relativistic, see Eqs.~(\ref{Delta-LDOS-AB}) and (\ref{DGviaI}) cases. The nonrelativistic
answer is written in terms of the function (\ref{I-final}) which is expressed as a combination of the
Whittaker functions in Eq.~(\ref{F-result}). The relativistic answer (\ref{DGviaI})
is expressed in terms of the same function  (\ref{I-final}), and a function (\ref{ID-def})
that describes the contribution of the $m=0$ solutions of the Dirac equation.

(ii) We show that in the vicinity of the vortex ($r \lesssim 0.2 l$) the relativistic LDOS is governed by the
function (\ref{ID-def}), so that in the limit $r \to 0$ the LDOS is given by Eq.~(\ref{LDOS-Dirac-small}).

(3) We obtained compact analytic expressions for the DOS perturbation by the Aharonov-Bohm flux. For the nonrelativistic
case, this is Eq.~(\ref{DOS-nonrel-vortex-const}), which in the clean limit reduces to the known\cite{Desbois1997NPB} result given by
Eq.~(\ref{DOS-nonrel-vortex-const-delta}). For the relativistic case the corresponding expressions for the DOS are
Eqs.~(\ref{DOS-rel-vortex-const-psi})  and (\ref{DOS-rel-vortex-const-delta}).

We hope that the obtained results will be useful both for the experimental STS studies of graphene
and for the theoretical studies of the interaction effects in an inhomogeneous magnetic field
similar to the recent work.\cite{Roy2011}  Among possible extensions of the considered problem we mention the
necessity to take into account a finite size of the vortex core, but this certainly demands more numerical
work, while in the present paper the main goal was to obtain some analytic results.

\section{Acknowledgments}

We thank V.P.~Gusynin for stimulating discussions.
This work was supported by the SCOPES Grant No. IZ73Z0\verb|_|128026 of
Swiss NSF, by the grant SIMTECH No. 246937 of the European FP7 program, and by
the Program of Fundamental Research of the Physics and Astronomy Division of
the National Academy of Sciences of Ukraine.
A.O.S. and S.G.S. were also supported by SFFR-RFBR Grant F40.2/108 ``Application of string theory
and field theory methods to nonlinear phenomena in low dimensional
systems.''

\appendix
\section{The calculation of the LDOS in the nonrelativistic case}
\label{sec:A}

Setting $\eta = 0$ in Eq.~(\ref{solution-nonrel}) one obtains
the solution of the Schr\"{o}dinger equation  for $B= \mbox{const}$ without vortex.
Substituting this solution in the LDOS defintion (\ref{LDOS-def}) and taking into account the widening of the LLs
(\ref{LL2Lorentzian}), we represent the LDOS as a double sum
\begin{equation}
\begin{split}
\label{DOS1}
N^{\mathrm{S}}_0(\mathbf{r},E,B)=& \frac{1}{\pi} \mbox{Im} \sum_{n=0}^\infty\sum_{m=-\infty}^\infty A^2_{n,m}y^{|m|}
e^{-y}[L_n^{|m|}(y)]^2 \\
& \times \frac{1}{E_{n,m}+ E_0 z},
\end{split}
\end{equation}
where in the second line, we introduced the dimensionless variable $z = - (E+ i \Gamma)/E_0$
with the characteristic energy $E_0$ defined below Eq.~(\ref{spectrum-nonrel}).
To calculate the sum in Eq.~(\ref{DOS1}), it is convenient to represent its last factor as an exponent
\begin{equation}
\label{frac2exp}
\begin{split}
& \frac{e^{-\delta(2n+|m|+m+1)}}{E_{n,m}+E_0z} \\
&= \frac{1}{E_0}\int_0^\infty d\beta e^{-(\beta+\delta)(2n+|m|+m+1)}e^{-\beta z}.
\end{split}
\end{equation}
Here, we also introduced the regularizing exponential factor with $\delta>0$, which makes the sum convergent and  will be
set to $0$ at the end. Then the LDOS acquires the form
\begin{widetext}
\begin{equation}
\label{DOS2}
N^{\mathrm{S}}_0(\mathbf{r},E,B)=\frac{M}{\pi^2 \hbar^2}
\mbox{Im} \left[\int_0^\infty  d\beta e^{-(\delta+\beta)}e^{-\beta z}
\sum_{m=-\infty}^\infty y^{|m|}e^{-y}e^{-(\beta+\delta)(|m|+m)}
\sum_{n=0}^\infty\frac{n!e^{-2(\beta+\delta)n}}{\Gamma(n+|m|+1)}[L_n^{|m|}(y)]^2\right].
\end{equation}
\end{widetext}
We operate with the representation (\ref{DOS2}) in the following way. First, we consider its analytic
continuation for $z>0$ and perform the calculation. Then to obtain the LDOS, we return to the
imaginary values $z \to - (E + i \Gamma)/E_0$ and evaluate the imaginary part.
Using Eq.~(10.12.20) from Ref.~\onlinecite{Bateman.book2}
\begin{widetext}
\begin{equation}
\label{Lauerre-sum}
\sum_{n=0}^\infty\frac{n!}{\Gamma(n+\alpha+1)}L_n^\alpha(x)L_n^\alpha(y)z^n=
(1-z)^{-1}\exp(-z\frac{x+y}{1-z})(xyz)^{-\frac{\alpha}{2}}I_\alpha\left(2\frac{\sqrt{xyz}}{1-z}\right), \quad |z|<1,
\end{equation}
\end{widetext}
where $I_\alpha$ is the modified Bessel function,
we find the sum over $n$ in Eq.~(\ref{DOS2})
\begin{widetext}
\begin{equation}
\label{DOS3}
N^{\mathrm{S}}_0(\mathbf{r},E,B)=\frac{M}{\pi^2 \hbar^2}
\mbox{Im}
\left[\int_0^\infty d\beta e^{-(\delta+\beta)}e^{-\beta z}
\frac{e^{-y\coth(\delta+\beta)}}{1-e^{-2(\delta+\beta)}}
\sum_{m=-\infty}^\infty e^{-(\delta+\beta)m}I_{|m|}\left(\frac{y}{\sinh(\delta+\beta)}\right)\right].
\end{equation}
\end{widetext}
The remaining summation over $m$ in Eq.~(\ref{DOS3}) can be done by using the property of the modified Bessel
function\cite{} $I_m(x)=I_{-m}(x)$, and that its generating function is\cite{Bateman.book2}
\begin{equation}
\label{generating-Bessel}
\sum_{m=-\infty}^{\infty} z^m I_m(x)= \exp{\left(\frac{x}{2}[z+1/z]\right)}.
\end{equation}
We obtain
\begin{equation}
\begin{split}
\label{DOS4}
& N^{\mathrm{S}}_0(E,B)=\\
& \frac{M}{(\pi\hbar)^2}\mbox{Im} \left[\int_0^\infty
d\beta\frac{e^{-(\delta+\beta)}e^{-\beta z}}{1-e^{-2(\delta+\beta})}\right].
\end{split}
\end{equation}
Notice that from the last expression one can explicitly observe that it does not depend on $y$, i.e.,
in a constant magnetic field the LDOS is position independent.
Introducing a new variable $x = 2 (\delta+\beta)$ we can rewrite the last expression as follows:
\begin{equation}
\begin{split}
\label{DOS5}
N^{\mathrm{S}}_0(E,B)=  - \frac{M}{2(\pi\hbar)^2} & \mbox{Im} \left[e^{\delta z}\int_{2\delta}^\infty
 dx\frac{e^{-x}-e^{-x(z+1)/2}}{1-e^{-x}} \right. \\
& \left.  -e^{\delta z}\int_{2\delta}^\infty
dx\frac{e^{-x}}{1-e^{-x}}\right].
\end{split}
\end{equation}
In the limit $\delta \to 0$, the second term of Eq.~(\ref{DOS5}) remains real irrespectively the
value of $z$, while the first term gives the integral representation of the digamma function:\cite{Bateman.book1}
\begin{equation}
\label{digamma-def}
\psi(z) = - \gamma + \int_{0}^{\infty} d t \frac{e^{-t} - e^{-tz}}{1-e^{-t}}, \quad \mbox{Re} (z) >0,
\end{equation}
where $\gamma$ is the Euler-Mascheroni constant. Thus we obtain
\begin{equation}
\label{LDOS-nonrel-const-z}
N^{\mathrm{S}}_0(E,B)=  - \frac{M}{2(\pi\hbar)^2} \mbox{Im} \left[\psi\left(\frac{z+1}{2}\right)\right],
\end{equation}
so that the final expression for the LDOS after the analytic continuation $z \to - 2(E+ i \Gamma)/(\hbar \omega_c)$
takes the form of Eq.~(\ref{LDOS-nonrel-const}). The oscillatory behavior of the
LDOS can be explicitly extracted from Eq.~(\ref{LDOS-nonrel-const-z}) [or Eq.~(\ref{LDOS-nonrel-const})] using the relationship
\begin{equation}
\label{psi-inv}
\psi(-z) = \psi(z) + \frac{1}{z} + \pi \cot(\pi z).
\end{equation}

Now we generalize these results for the case when the vortex is present. Repeating the steps that led us from Eq.~(\ref{DOS1}) to
Eq.~(\ref{DOS3}), we obtain
\begin{widetext}
\begin{equation}
\label{DOSv1}
N^{\mathrm{S}}_0(\mathbf{r},E,B)=\frac{M}{\pi^2 \hbar^2}
\mbox{Im} \left[\int_0^\infty d\beta e^{-(\delta+\beta)}e^{-\beta z}
\frac{e^{-y\coth(\delta+\beta)}}{1-e^{-2(\delta+\beta)}}
\sum_{m=-\infty}^\infty e^{-(\delta+\beta)(m+\eta)}
I_{|m+\eta|}\left(\frac{y}{\sinh(\delta+\beta)}\right)\right].
\end{equation}
\end{widetext}
The sum over $m$ in Eq.~(\ref{DOSv1}) is calculated in Appendix~\ref{sec:B}. Using Eq.~(\ref{sum-m-final})
we obtain
\begin{equation}
\label{sum}
\begin{split}
& \sum_{m=-\infty}^\infty e^{-(\delta+\beta)(m+\eta)}
I_{|m+\eta|}\left(\frac{y}{\sinh(\delta+\beta)}\right)= \\
& e^{y\coth(\delta+\beta)}-\frac{\sin\pi\eta}{\pi}\\
& \times \int_{-\infty}^\infty d\omega e^{-y\cosh \omega/\sinh(\delta+\beta)}
\frac{e^{-\eta(\delta+\beta+\omega)}}{1+e^{-(\delta+\beta+\omega)}}.
\end{split}
\end{equation}
The first term on the right-hand side of the last equation corresponds to the LDOS without the vortex, which
was considered above, so that we can concentrate on the second term. Substituting it in Eq.~(\ref{DOSv1}), we arrive at
Eq.~(\ref{delta1-nonrel-LDOS}) for $\Delta N_\eta^{\mathrm{S}}(\mathbf{r},E,B)$.

\section{The calculation of the sum over the azimuthal quantum number}
\label{sec:B}

The sum over the azimuthal quantum number
\begin{equation}
\label{sum-m-def}
\Sigma(\eta)=\sum_{m=-\infty}^\infty e^{-\beta(m+\eta)}I_{|m+\eta|}(x).
\end{equation}
can be found using the method described in Ref.~\onlinecite{Marino1982NPB}.
Using the integral representation of the modified Bessel function\cite{Gradshteyn:book}
\begin{equation}
I_\nu (z) = \frac{1}{2\pi i}\int_C e^{z \cosh \omega - \nu \omega} d \omega,
\end{equation}
where $C$ is a complex path beginning at $-i \pi + \infty$ and ending at $i \pi + \infty$,
we obtain
\begin{equation}
\begin{split}
\Sigma & (\eta)=\frac{1}{2\pi i}\int_C d\omega e^{x\cosh\omega}\\
& \times \left[\sum_{m=0}^\infty e^{-(\beta+\omega)(m+\eta)}+
\sum_{m=1}^\infty e^{-(\omega-\beta)(m-\eta)}\right].
\end{split}
\end{equation}
Choosing the contour $C$ to lie in such a way that the condition $\mbox{Re} \omega > \beta$ is satisfied,
the series can be made convergent, so that
\begin{equation}
\Sigma(\eta)=\frac{1}{2\pi i}\int_C d\omega e^{x\cosh\omega}
\left[\frac{e^{-(\beta+\omega)\eta}}{1-e^{-(\beta+\omega)}}+
\frac{e^{(\omega-\beta)\eta}}{e^{(\omega-\beta)}-1}\right].
\end{equation}
Now changing the variable $\omega \to - \omega$ in the second integral we can write
\begin{equation}
\Sigma(\eta)=\frac{1}{2\pi i}\left[\int_C+\int_{C'}\right] d\omega e^{x\cosh\omega}\frac{e^{-(\beta+\omega)\eta}}{1-e^{-(\beta+\omega)}},
\end{equation}
where $C'$ is the contour symmetric to the contour $C$ with respect to the origin of coordinates.
Joining the contours $C$ and $C'$ leads to an $\omega$ integral of the form
\begin{equation}
\int_C d\omega+\int_{C'} d\omega=\int_{-\infty+i\pi}^{\infty+i\pi} d\omega +\int_{\infty-i\pi}^{-\infty-i\pi} d\omega
+\oint_{C''} d\omega,
\end{equation}
where $C''$ is a rectangle with a length larger than $2 \beta$ and  width $2 \pi i$, centered in the origin,
traversed anticlockwise.
Inside the contour $C''$, the integrand has only one pole at $\omega_0=-\beta$, so that this integral does not depend
on $\eta$ and corresponds to $\Sigma(0)$. Therefore, we arrive at the final representation for the sum (\ref{sum-m-def}):
\begin{equation}
\label{sum-m-final}
\Sigma(\eta)=-\frac{\sin\pi\eta}{\pi}\int_{-\infty}^\infty d\omega e^{-x\cosh\omega}\frac{e^{-(\beta+\omega)\eta}}{1+e^{-(\beta+\omega)}}+\Sigma(0),
\end{equation}
where
\begin{equation}
\label{Sigma-0}
\Sigma(0)=e^{x\cosh\beta}.
\end{equation}
Finally, note that one can reproduce the value $\Sigma(0)$ from Eq.~(\ref{Sigma-0}) using
Eq.~(\ref{generating-Bessel}), which for $z = e^{-\beta}$, reduces to the sum (\ref{sum-m-def})
with $\eta=0$.

\section{The calculation of the density of states in the nonrelativistic case}
\label{sec:C}

Substituting Eq.~(\ref{delta1-nonrel-LDOS}) in the definition (\ref{DOS-nonrel}) and integrating over the spatial coordinates,
we obtain
\begin{equation}
\label{delta-DOS-nonrel1}
\begin{split}
 \Delta & N_\eta^{\mathrm{S}}(E,B)=-\sin\pi\eta\frac{Ml^2}{2(\pi\hbar)^2}
\mbox{Im} \left[\int_0^\infty d\beta\int_{-\infty}^\infty d\upsilon \right.
\\ & \left.
\frac{e^{-\beta z}}{\cosh (\upsilon/2)\cosh(\beta+ \delta-\upsilon/2)}
\frac{e^{-\eta \upsilon}}{1+e^{-\upsilon}}\right],
\end{split}
\end{equation}
where we introduced the new variable $\upsilon = \omega + \beta +\delta$.
This double integral can be rewritten using the new variables
$t=e^{-2\beta},\, x=e^\upsilon$ as follows:
\begin{equation}
\label{delta-DOS-nonrel2}
\begin{split}
& \Delta  N_\eta^{\mathrm{S}}(E,B)=-\sin\pi\eta\frac{Ml^2e^{-\delta}}{(\pi\hbar)^2} \\
& \times \mbox{Im} \left[\int_0^1 dt t^{(z-1)/2} \int_0^\infty \frac{dx x^{1-\eta}}{(1+x)^2(1+te^{-2\delta}x)}\right],
\end{split}
\end{equation}
where the second integral can be calculated using the residue theory
\begin{equation}
\begin{split}
& \int_0^\infty \frac{dx x^{1-\eta}}{(1+x)^2(1+te^{-2\delta}x)} \\
& =\frac{\pi}{\sin\pi\eta}
\frac{1-\eta+\eta e^{-2\delta}t-e^{-2\eta\delta}t^\eta}{(1-e^{-2\delta}t)^2}.
\end{split}
\end{equation}
Then the remaining integral is expressed via the hypergeometric function
\begin{equation}
\label{int-hypergeometric}
\begin{split}
&\int_0^1 dt t^{(z-1)/2}\frac{1-\eta+\eta e^{-2\delta}t-e^{-2\eta\delta}t^\eta}{(1-e^{-2\delta}t)^2}  = \frac{1-e^{-2\delta\eta}}{1-e^{-2\delta}} \\
&- (z+2\eta-1)\left[ \frac{1}{1+z}\,_2F_1 \left(1,\frac{1+z}{2}; \frac{3+z}{2};e^{-2\delta} \right) \right.\\
& \left. -\frac{e^{-2\delta\eta}}{1+z+2\eta}\, _2F_1 \left(1, \frac{1+z}{2}+\eta; \frac{3+z}{2}+\eta;e^{-2\delta} \right) \right].
\end{split}
\end{equation}
Now we use the series representation of hypergeometric functions in Eq.~(\ref{int-hypergeometric}):
\begin{equation}
\begin{split}
& \frac{e^{-2\delta\eta} }{z+2\eta+1}\, _2F_1 \left( 1, \frac{1+z}{2}+\eta,\frac{3+z}{2}+\eta,e^{-2\delta} \right)=\\
& = \sum_{n=0}^\infty \frac{e^{-2\delta(n+\eta)}}{z+1+2\eta+2n} \\
&=e^{\delta(z+1)}\sum_{n=0}^\infty\int_\delta^\infty dxe^{-x(2n+2\eta+z+1)}\\
&=e^{\delta(z+1)}\int_\delta^\infty dx\frac{e^{-x(z+1+2\eta)}}{1-e^{-2x}},
\end{split}
\end{equation}
where the first one in Eq.~(\ref{int-hypergeometric}) is recovered for $\eta=0$.
We observe that the presence of finite $\delta>0$ makes the hypergeometric series well defined, but at
the end of the calculation the limit $\delta\to 0$ can already be taken. Then
taking into account the integral representation of  the digamma function (\ref{digamma-def}) [similarly to Eq.~(\ref{DOS5})]
one can express the DOS (\ref{delta-DOS-nonrel2}) in the following simple form:
\begin{equation}
\label{DOS-nonrel-vortex-const-z}
\begin{split}
&\Delta  N_\eta^{\mathrm{S}}(E,B)=\frac{Ml^2}{2\pi\hbar^2} \mbox{Im} \bigg\{(z+2\eta-1) \\
& \left. \times \left[ \psi \left( \frac{z+1}{2}+\eta \right)-\psi \left(\frac{z+1}{2}\right) \right] \right\},
\end{split}
\end{equation}
which after the analytic continuation $z \to -2(E+i \Gamma)/(\hbar \omega_c)$
takes the final form (\ref{DOS-nonrel-vortex-const}).

\section{Calculation of the function $I(y,z,\eta)$}
\label{sec:D}

As in Ref.~\onlinecite{Slobodeniuk2010PRB} we observe that it is simpler to calculate integrals with the derivative
$d I(y,z,\eta)/d y$ representing the function $I(y,z,\eta)$ in the form:
\begin{equation}
\label{I-integral}
I(y,z,\eta)=-\int_y^\infty\frac{dI(Q,z,\eta)}{dQ},
\end{equation}
where we used that $I(\infty,z,\eta)=0$. The derivative $dI(Q,z,\eta)/dQ$ contains two terms;
\begin{equation}
\label{I1+I2}
\frac{dI(Q,z,\eta)}{dQ}=\frac{dI_1(Q,z,\eta)}{dQ}+\frac{dI_2(Q,z,\eta)}{dQ},
\end{equation}
where
\begin{equation}
\begin{split}
&\frac{dI_1}{dQ}=-\frac12\int_0^\infty d\beta e^{-\beta (z+\eta)}
\frac{e^{-Q\coth\beta}}{\sinh^2\beta} \\
& \times \int_{-\infty}^\infty d\omega e^{-Q\cosh\omega/\sinh\beta}
e^{-(\eta-1)\omega},\\
& \frac{dI_2}{dQ}=-\frac12\int_0^\infty d\beta e^{-\beta (z+\eta-1)}
\frac{e^{-Q\coth\beta}}{\sinh^2\beta} \\
& \times \int_{-\infty}^\infty d\omega e^{-Q\cosh\omega/\sinh\beta}
e^{-\eta\omega}.
\end{split}
\end{equation}
Using the integral representation of the MacDonald function $K_\nu(x)$ (see Ref.~\onlinecite{Bateman.book2})
\begin{equation}
K_\nu(x)=\frac12\int_{-\infty}^\infty e^{-x\cosh\omega-\nu \omega}d\omega,
\end{equation}
we obtain
\begin{equation}
\frac{dI_1}{dQ}=-\int_0^\infty d\beta e^{-\beta (z+\eta)}
 \frac{e^{-Q\coth\beta}}{\sinh^2\beta}
K_{1-\eta}(Q/\sinh\beta)
\end{equation}
and
\begin{equation}
\frac{dI_2}{dQ}=-\int_0^\infty d\beta e^{-\beta (z+\eta-1)}
 \frac{e^{-Q\coth\beta}}{\sinh^2\beta}
K_\eta(Q/\sinh\beta).
\end{equation}
Now introducing a new variable $t$ via $e^{-2\beta}=t/(1+t)$, we get
\begin{equation}
\label{I1-der}
\begin{split}
\frac{dI_1}{dQ}=-2e^{-Q}\int_0^\infty
dt & t^{(z+\eta)/2}(1+t)^{-(z+\eta)/2}e^{-2Qt} \\
& \times K_{1-\eta}[2Q\sqrt{t(1+t)}]
\end{split}
\end{equation}
and
\begin{equation}
\label{I2-der}
\begin{split}
\frac{dI_2}{dQ}=-2e^{-Q}\int_0^\infty
dt & t^{(z+\eta-1)/2}(1+t)^{-(z+\eta-1)/2}e^{-2Qt} \\
& \times K_\eta[2Q\sqrt{t(1+t)}].
\end{split}
\end{equation}
To integrate over $t$ in Eqs.~(\ref{I1-der}) and (\ref{I2-der}), we use the integral (2.16.10.2)
from Ref.~\onlinecite{Prudnikov.vol2}:
\begin{equation}
\label{Prud-int}
\begin{split}
\int_0^\infty & dx \frac{x^{\rho-1}}{(x+z)^\rho} e^{-px} K_\nu(c\sqrt{x^2+xz})=\\
& \frac{1}{cz} \Gamma\left( \rho+\frac{\nu}{2} \right) \Gamma \left(\rho-\frac{\nu}{2} \right) e^{pz/2} \\
& \times W_{1/2-\rho,\nu/2}\left(z_+ /2\right) W_{1/2-\rho,\nu/2} \left(z_- /2 \right), \\
& z_\pm = z (p \pm \sqrt{p^2-c^2}), \\
& \mbox{Re} (p+c) >0, |\mbox{arg z}| < \pi, \quad 2 \mbox{Re} (\rho) > |\mbox{Re} (\nu)|,
\end{split}
\end{equation}
where $W_{\lambda,\mu} (z)$ is the Whittaker function. To adapt Eq.~(\ref{Prud-int}) to the form of Eqs.~(\ref{I1-der}) and (\ref{I2-der}), we have to set $z=1$, differentiate the result over $p$, and then take the limit $p \to c$. This gives
\begin{equation}
\begin{split}
& \int_0^\infty  dx \frac{x^\rho}{(x+1)^\rho}e^{-cx}K_\nu[c\sqrt{x(x+1)}]= \\
& -\frac{1}{2}\Gamma\left(\rho+\frac{\nu}{2}\right)\Gamma\left(\rho-\frac{\nu}{2}\right) e^{c/2}
G_{1/2-\rho,\nu/2} \left( \frac{c}{2} \right),
\end{split}
\end{equation}
where the function $G_{\lambda,\mu} (Q)$ is defined as follows:
\begin{equation}
\begin{split}
G_{\lambda,\mu}  (Q)&=\frac{1}{2Q}W^2_{\lambda,\mu}(Q)+\frac1QW_{\lambda,\mu}(Q)W'_{\lambda,\mu}(Q) \\
& + W''_{\lambda,\mu}(Q)W_{\lambda,\mu}(Q)-W'^2_{\lambda,\mu}(Q).
\end{split}
\end{equation}
Accordingly, we obtain that
\begin{equation}
\label{I1viaG}
\begin{split}
\frac{dI_1}{dQ}= \Gamma\left(\frac{z+1}{2}\right)\Gamma\left(\frac{z+2\eta-1}{2}\right) G_{(1-z-\eta)/2,(1-\eta)/2}(Q)
\end{split}
\end{equation}
and
\begin{equation}
\label{I2viaG}
\begin{split}
\frac{dI_2}{dQ}= \Gamma\left(\frac{z-1}{2}\right)\Gamma\left(\frac{z+2\eta-1}{2}\right) G_{(2-z-\eta)/2,\eta/2}(Q).
\end{split}
\end{equation}
Now using the differential equation
\begin{equation}
\label{Whittaker-dif}
W''_{\lambda,\mu}(z)+\left(-\frac14+\frac{\lambda}{z}+
\frac{1/4-\mu^2}{z^2}\right)W_{\lambda,\mu}(z)=0
\end{equation}
and the recursion formula
\begin{equation}
\label{Whitaker-recurison}
\begin{split}
z\frac{d}{dz}W_{\lambda,\mu}(z) & =\left(\lambda- \frac{z}{2} \right)W_{\lambda,\mu}(z) \\
&- \left[ \mu^2-\left(\lambda- \frac{1}{2} \right)^2 \right]W_{\lambda-1,\mu}(z)
\end{split}
\end{equation}
for the Whittaker function,\cite{Gradshteyn:book} one can transform $G_{\lambda,\mu} (Q)$
to the form
\begin{equation}
\begin{split}
G_{\lambda,\mu} & (Q)=\frac{\mu^2+(\lambda-1/2)^2}{Q^2}W^2_{\lambda,\mu}(Q)\\
&- \frac{[\mu^2-(\lambda-1/2)^2]^2}{Q^2}W^2_{\lambda-1,\mu}(Q)\\
&- \frac{\mu^2-(\lambda-1/2)^2}{Q}W_{\lambda,\mu}(Q)W_{\lambda-1,\mu}(Q)\\
-\frac{2\lambda-1}{2Q} & W^2_{\lambda,\mu}(Q)-
(\lambda-1/2)\left(\frac{W^{2}_{\lambda,\mu}(Q)}{Q}\right)'.
\end{split}
\end{equation}
To obtain the function $F_{\lambda,\mu} (y) = -\int_y^\infty d Q G_{\lambda,\mu} (Q)$, we employ the relationships:
\begin{equation}
\begin{split}
& \int \frac{dQ}{Q}W_{\lambda,\mu}(Q)W_{\rho,\mu}(Q)=\\
&\frac{1}{\rho-\lambda}[W'_{\lambda,\mu}(Q)W_{\rho,\mu}(Q)-
W'_{\rho,\mu}(Q)W_{\lambda,\mu}(Q)],\\
& \int \frac{dQ}{Q}W_{\lambda,\mu}(Q)W_{\lambda,\mu}(Q)=\\
& W'_{\lambda,\mu}(Q)\partial_\lambda W_{\lambda,\mu}(Q)-
\partial_\lambda W'_{\lambda,\mu}(Q)W_{\lambda,\mu}(Q),\\
&\int\frac{dQ}{Q^2}W_{\lambda,\nu}(Q)W_{\lambda,\nu}(Q) =\\
& \frac{1}{2\nu}[\partial_\nu W'_{\lambda,\nu}(Q)W_{\lambda,\nu}(Q)-W'_{\lambda,\nu}(Q)
\partial_\nu W_{\lambda,\nu}(Q)],
\end{split}
\end{equation}
which follow from the differential equation (\ref{Whittaker-dif}) for the Whittaker function. Then using
the recursion formula (\ref{Whitaker-recurison}), we arrive at the following result:
\begin{widetext}
\begin{equation}
\label{F-result}
\begin{split}
F_{\lambda,\mu}(y)=&\frac{\mu^2+(\lambda-1/2)^2}{2\mu y} [W_{\lambda+1,\mu}(y)
\partial_\mu W_{\lambda,\mu}(y)-W_{\lambda,\mu}(y)
\partial_\mu W_{\lambda+1,\mu}(y)]\\-&\frac{[\mu^2-(\lambda-1/2)^2]^2}
{2\mu y}[W_{\lambda,\mu}(y)\partial_\mu W_{\lambda-1,\mu}(y)-W_{\lambda-1,\mu}(y)
\partial_\mu W_{\lambda,\mu}(y)]\\+&
\frac{\mu^2-(\lambda-1/2)^2}{y}[W_{\lambda,\mu}^2(y)-W_{\lambda-1,\mu}(y)W_{\lambda,\mu}(y)-
W_{\lambda-1,\mu}(y)W_{\lambda+1,\mu}(y)]\\-&
\frac{2\lambda-1}{2 y}[2W_{\lambda,\mu}^2(y)-W_{\lambda+1,\mu}(y)\partial_\lambda W_{\lambda,\mu}(y)+
W_{\lambda,\mu}(y)\partial_\lambda W_{\lambda+1,\mu}(y)].
\end{split}
\end{equation}
\end{widetext}
The integral of each term in Eq.~(\ref{I1+I2}) is expressed
via  $F_{\lambda,\mu}(y)$ with the prefactors given by Eqs.~(\ref{I1viaG}) and (\ref{I2viaG}), so that we arrive at the
final expression (\ref{I-final}) for the function $I(y,z,\eta)$ which was defined in Eq.~(\ref{I-integral}).

To complete our analysis, we consider the asymptotic of the  $\mbox{Im}I(y,z\to -(E+i\Gamma)/E_0,\eta)$
in the limits $y \to 0$ and $y \to \infty$. To do this we use the following representations of
the Whittaker function:
\begin{equation}
\label{Whittaker-small}
\begin{split}
& W_{\lambda,\mu}  (y) \approx  y^{1/2-\mu}\frac{\Gamma(2\mu)}{\Gamma(1/2+\mu-\lambda)} + O(y^{3/2-\mu})\\
& + y^{1/2+\mu}\frac{\Gamma(-2\mu)}{\Gamma(1/2-\mu-\lambda)} + O(y^{3/2+\mu}), \quad y \to 0,
\end{split}
\end{equation}
and
\begin{equation}
\label{Whittaker-large}
W_{\lambda,\mu}  (y) \approx e^{-y/2}y^\lambda\left[1 + O(1/y) \right], \quad y \to \infty.
\end{equation}
Substituting Eq.~(\ref{Whittaker-small}) in Eq.~(\ref{I-final}) and omitting all real terms, which
will not contribute to  $\mbox{Im} I$, we obtain
\begin{equation}
\begin{split}
& \mbox{Im} I(y \to 0,z,\eta) \\
& \approx -
\frac{\pi}{\sin \pi \eta}\mbox{Im} \left[\psi\left(\frac{z-3}{2} \right) + \frac{4(z-2)}{(z-1)(z-3)} \right].
\end{split}
\end{equation}
Now using the property of the digamma function
\begin{equation}
\label{psi-recurrent}
\psi(z) = \psi(z+1) - \frac{1}{z},
\end{equation}
we arrive at the result
\begin{equation}
\label{I-small-y}
\mbox{Im} I(y = 0,z,\eta) = - \frac{\pi}{\sin \pi \eta} \mbox{Im} \psi\left(\frac{1+z}{2} \right).
\end{equation}
One can reproduce the same result directly from Eq.~(\ref{I-def}).
Indeed, setting $y=0$ in Eq.~(\ref{I-def}) we have
\begin{equation}
I(y=0,z,\eta) =
\int_0^\infty d\beta \frac{e^{-\beta z}}{\sinh\beta}
\int_{-\infty}^\infty d\omega \frac{e^{-\eta\omega}}{1+e^{-\omega}}.
\end{equation}
Now the integral over $\omega$ is elementary and after replacing $2 \beta \to \beta$,
we obtain
\begin{equation}
I(y=0,z,\eta) = \frac{\pi}{\sin \pi \eta}  \int_0^\infty d\beta\frac{e^{-\beta (z+1)/2}}{1-e^{-\beta}}.
\end{equation}
Recognizing in this integral the imaginary part of the digamma function
(\ref{digamma-def}), we again arrive at Eq.~(\ref{I-small-y}).
The next order corrections to $\mbox{Im} I(y \to 0,z,\eta) $ can be obtained by expanding
the function $d I/dQ$ at $Q=0$, integrating the result over $Q$, and using the $y=0$ result (\ref{I-small-y}).
For $y \to 0$ the expansion  contains terms $\sim y^{1-\eta}$ and $y^\eta$ with  prefactors that make the
resulting approximate expression for the LDOS divergent at $\eta =0,1$.

Substituting the asymptotic (\ref{Whittaker-large}) in Eq.~(\ref{I-final}), we obtain that for $y \to \infty$
\begin{equation}
\label{F-large-y}
F_{\lambda,\mu}(y)\sim e^{-y}y^{2\lambda-3}\left(\frac14+\lambda(\lambda-1)-\mu^2\right).
\end{equation}

\section{Solution of the Dirac equation}
\label{sec:E}

A positive energy solution of the time-dependent Dirac equation has a form
$\Psi(t, \mathbf{r}) = \exp(-i Et/\hbar)\Psi(\mathbf{r})$, where the components of
a two-component spinor
\begin{equation}
\label{psi-def-appendix}
\Psi(\mathbf{r},\zeta) = \left[\begin{array}{cc} \psi_1({\mathbf r,\zeta})\\
i\psi_2({\mathbf r,\zeta})  \end{array}\right]
\end{equation}
satisfy   the following equations [compare with Appendix~A of Ref.~\onlinecite{Slobodeniuk2010PRB}]:
\begin{equation}
\label{components-polar}
\begin{split}
& (E-\Delta)\psi_1(\mathbf{r},\zeta)\\
&-\hbar v_Fe^{-i\zeta\varphi}\left(\frac{\partial}
{\partial r}-\frac{i\zeta}{r}\frac{\partial}{\partial\varphi}+
\frac{e\zeta A_\varphi}{\hbar c}\right)\psi_2(\mathbf{r},\zeta)=  0 , \\
& \hbar v_Fe^{i\zeta\varphi}\left(\frac{\partial}
{\partial r}+\frac{i\zeta}{r}\frac{\partial}{\partial\varphi}-
\frac{e\zeta A_\varphi}{\hbar c}\right)\psi_1(\mathbf{r},\zeta)\\
&+(E+\Delta)\psi_2(\mathbf{r},\zeta)=  0.
\end{split}
\end{equation}
The vector potential $A_\varphi(r)$ in Eq.~(\ref{components-polar}) is given by
Eq.~(\ref{reg-potential}). From now on, we consider the specific case $\zeta =1$
(omitting the label $\zeta$ in the wave functions)
and seek for a solution of Eq.~(\ref{components-polar}) in the following form:
\begin{equation}
\psi_1(\mathbf{r})=e^{i(m-1)\varphi}\psi_1(r),\qquad \psi_2(\mathbf{r})=e^{im\varphi}\psi_2(r).
\end{equation}
Then the radial components of the spinor $\psi_1(r)$ and $\psi_2(r)$ satisfy the following
system of equations
\begin{equation}
\label{radial-system}
\begin{split}
& \psi_1(r)=\frac{\hbar v_F}{E-\Delta}\left(\frac{d}{dr}+
\frac{m+\eta\theta(r-R)}{r}+
\frac{r}{2l^2}\right)\psi_2(r),\\
& \psi_2(r)=-\frac{\hbar v_F}{E+\Delta}\left(\frac{d}{dr}-
\frac{m+\eta\theta(r-R)-1}{r}-
\frac{r}{2l^2}\right)\psi_1(r).
\end{split}
\end{equation}
Introducing the dimensionless variable $y = r^2/(2l^2)$ and denoting the $\rho \equiv R^2/(2 l^2)$, we
rewrite the system (\ref{radial-system}) for $y\in[0,\rho]$
\begin{equation}
\label{system:r<R}
\begin{split}
& \psi_1(y)=\frac{\hbar v_F \sqrt{2}}{(E-\Delta)l}\sqrt{y}
\left(\frac{d}{dy}+ \frac{m}{2y}+ \frac{1}{2}\right)\psi_2(y), \\
& \psi_2(y)=-\frac{\hbar v_F \sqrt{2}}{(E+\Delta)l}\sqrt{y}
\left(\frac{d}{dy}- \frac{m-1}{y}- \frac{1}{2}\right)\psi_1(y) .
\end{split}
\end{equation}
Since there is no Aharonov-Bohm field for $y < \rho$, the problem in this domain is identical
to that of the Appendix~D in Ref.~\onlinecite{Gusynin2006PRB}.
For $y\in[\rho,\infty[$, the system (\ref{radial-system}) acquires the form
\begin{subequations}
\label{system:r>R}
\begin{align}
& \psi_1(y)=\frac{\hbar v_F \sqrt{2}}{(E-\Delta)l}\sqrt{y}
\left(\frac{d}{dy}+ \frac{m+\eta}{2y}+ \frac{1}{2}\right)\psi_2(y),\\
& \psi_2(y)=-\frac{\hbar v_F \sqrt{2}}{(E+\Delta)l}\sqrt{y}
\left(\frac{d}{dy}- \frac{m+\eta-1}{2y}- \frac{1}{2}\right)\psi_1(y). \label{psi2viapsi1}
\end{align}
\end{subequations}
The matching conditions (\ref{continuity}) and (\ref{discontinuity}) take the form
\begin{equation}
\label{psi1-match}
\begin{split}
& \psi_1(\rho-0)=\psi_1(\rho+0),\\
& \psi_1'(\rho+0)-\psi_1'(\rho-0)=\frac{\eta}{2\rho}\psi_1(\rho),
\end{split}
\end{equation}
and
\begin{equation}
\begin{split}
& \psi_2(\rho-0)=\psi_2(\rho+0),\\
& \psi_2'(\rho+0)-\psi_2'(\rho-0)=-\frac{\eta}{2\rho}\psi_2(\rho),
\end{split}
\end{equation}
where the derivative is taken over $y$.
One can obtain from the system (\ref{system:r<R}) that for $y\in[0,\rho]$
the spinor components
satisfy the following second-order differential
equations:
\begin{subequations}
\label{2nd-order:r<R}
\begin{align}
& \left\{\frac{d^2}{dy^2}+\frac{1}{y}\frac{d}{dy} - \frac{1}{4} - \frac{(m-1)^2}{4y^2}  +
\frac{\lambda-m}{2y} \right\}\psi_1(y)=0, \label{psi1:r<R} \\
& \left\{\frac{d^2}{dy^2}+\frac{1}{y}\frac{d}{dy} - \frac{1}{4} - \frac{m^2}{4y^2}
+\frac{\lambda-m+1}{2y}\right\}\psi_2(y)=0, \label{psi2:r<R}
\end{align}
\end{subequations}
where we introduced $\lambda=(E^2-\Delta^2)l^2/(\hbar v_F)^2$.
The  second order differential equations for the domain
$y\in[\rho,\infty[$ corresponding to the system (\ref{system:r>R}) can be obtained from
Eq.~(\ref{2nd-order:r<R}) by replacing $m \to m + \eta$:
\begin{subequations}
\label{2nd-order:r>R}
\begin{align}
& \left\{\frac{d^2}{dy^2}+\frac{1}{y}\frac{d}{dy} - \frac{1}{4} - \frac{(m+\eta-1)^2}{4y^2}
+ \frac{\lambda-m-\eta}{2y} \right\}\psi_1(y)=0,  \label{psi1:r>R} \\
& \left\{\frac{d^2}{dy^2}+\frac{1}{y}\frac{d}{dy} - \frac{1}{4} - \frac{(m+\eta)^2}{4y^2}
+\frac{\lambda-m-\eta+1}{2y}\right\}\psi_2(y)=0.
\end{align}
\end{subequations}
The equations can be reduced to the equations
for the degenerate hypergeometric function (see Eq.~(6.3.1) of Ref.~\onlinecite{Bateman.book1})
and the solutions of Eqs.~(\ref{psi1:r<R}) and (\ref{psi1:r>R}) are given, respectively, by
\begin{widetext}
\begin{subequations}
\label{solution-psi1}
\begin{align}
 \psi_1(y)=&C_m y^{|m-1|/2}e^{-y/2}  \Phi \left( \frac{|m-1|+m+1-\lambda}{2},1+|m-1|;y \right), \qquad r < R
\label{solution-psi1:r<R} \\
\psi_1 (y)=&  A_m y^{|m+\eta-1|/2} e^{-y/2}\Phi \left( \frac{a_+ -\lambda}{2},1+|m+\eta-1|;y \right) \nonumber \\
 + &B_m y^{-|m+\eta-1|/2} e^{-y/2} \Phi \left( \frac{a_- -\lambda}{2},1-|m+\eta-1|;y \right), \qquad r >R,
\label{solution-psi1:r>R}
\end{align}
\end{subequations}
\end{widetext}
where $a_{\pm} \equiv m+\eta+1 \pm |m+\eta-1|$,
$A_m$, $B_m$, and $C_m$ are constants, and $\Phi(a,c;z)$ is the confluent hypergeometric  function.
The solution (\ref{solution-psi1:r<R}) contains only one term due to the condition of square integrability
and the absence of the Aharonov-Bohm field for $r < R$. Writing the solution (\ref{solution-psi1:r>R}) we used that
for noninteger $c$ the solution of Eq.~(\ref{psi1:r>R}) can be expressed via $\Phi(a,c;z)$ and $z^{1-c} \Phi(a-c+1,2-c;z)$.
The coefficients $A_m$, $B_m$, and $C_m$ to be found from the matching conditions (\ref{psi1-match}).
The consideration of the limit $R \to 0$ ($\rho \to 0$) greatly simplifies the calculation, because one
can expand the solutions to the linear in $\rho$ terms. Then one finds that
\begin{equation}
\label{solution-psi1-Phi}
\begin{split}
& \psi_1(y)=A_m y^{|m+\eta-1|/2}e^{-y/2} \\
& \times \Phi \left( \frac{|m+\eta-1|+m+\eta+1-\lambda}{2},1+|m+\eta-1|;y \right).
\end{split}
\end{equation}
Since $\Phi(a,c;z)$ behaves as $e^y$ at large $y$ unless $a= -n$ with $n= 0,1,2,\ldots$,
in order to have the square integrable solutions, the value $\lambda$ should be equal to the eigenvalue
$\lambda_{m,n}$ defined by Eq.~(\ref{spectrum-rel}).
In this case $\Phi$ function is reduced to the generalized Laguerre polynomials [see Eq.~(6.9.2.36) of
Ref.~\onlinecite{Bateman.book1}]
\begin{equation}
L_n^\alpha(y) = \frac{\Gamma(\alpha+n+1)}{\Gamma(\alpha+1) n!} \Phi(-n,\alpha+1,y).
\end{equation}
Introducing the functions
\begin{equation}
\label{J-def}
J^n_\nu(x)=\left( \frac{\Gamma(n+1)}{\Gamma(n+\nu+1)} \right)^{1/2}
e^{-x/2}x^{\nu/2}L_n^\nu(x),
\end{equation}
one can rewrite the solution (\ref{solution-psi1-Phi}) in a  more compact form
\begin{equation}
\psi_1(y)=A_m J^n_{|m+\eta -1|}(y).
\end{equation}
The definition (\ref{J-def}) generalizes  the functions considered in Ref.~\onlinecite{Melrose1983AJP} for the case of
the noninteger $\nu >-1$. These functions satisfy the following orthogonality condition
\begin{equation}
\int\limits_0^\infty dx J_\nu^n(x)J_\nu^{n^\prime}(x)=\delta_{nn^\prime}.
\end{equation}
Then having $\psi_1(y)$ one can find $\psi_2(y)$ from Eq.~(\ref{psi2viapsi1}) using the recursion formulas\cite{Melrose1983AJP}
\begin{equation}
\begin{split}
& (x+\nu)J^n_\nu(x)=\\
& [x(n+\nu)]^{\frac{1}{2}}J^n_{\nu-1}(x)+[x(n+\nu+1)]^{\frac{1}{2}} J^n_{\nu+1}(x),\\
& 2x(d/dx)J^n_\nu(x)=\\
&[x(n+\nu)]^{\frac{1}{2}}J^n_{\nu-1}(x)-[x(n+\nu+1)]^{\frac{1}{2}}
J^n_{\nu+1}(x).
\end{split}
\end{equation}
Then demanding that the spinors obey the normalization condition
\begin{equation}
\int_{0}^{2 \pi} d \varphi \int_{0}^{\infty} r d r \Psi^\dagger_{n^\prime m^\prime}(\mathbf{r},\zeta)
\Psi_{n m}(\mathbf{r},\zeta) = \delta_{n,n^\prime} \delta_{m,m^\prime}
\end{equation}
we obtain the solutions (\ref{Dirac:sol-m>0}), (\ref{Dirac:sol-m=0}), and (\ref{Dirac:sol-m<0}) for $n>0$.
The zero-mode solutions have to be considered separately. Analyzing the initial system
(\ref{system:r>R}), one finds that the only allowed solution is the negative energy
$E=-\Delta$, $m\leq 0$ with $\psi_1(r) =0$. The corresponding spinor is given by Eq.~(\ref{Dirac:sol-n=0}).
One can verify that for $\eta=0$ these solutions transform up to the phase factors to the solutions
obtained in Ref.~\onlinecite{Gusynin2006PRB}.
To show this, one should relabel the quantum numbers $n+m \to n$ for $m \geq 1$ and $n+1 \to n$ for $m \leq 0$
and use the property $J^n_\nu(y)=(-1)^\nu J^{n+\nu}_{-\nu}(y)$,  which is valid only when $J^n_\nu(y)$ is defined
for the integer values of $\nu$ as done in Ref.~\onlinecite{Melrose1983AJP}. After this relabeling is made, the spectrum
(\ref{spectrum-rel}) acquires a conventional form dependent only on the LL index which for $\Delta =0$
reduces to Eq.~(\ref{Dirac-LL}).

\end{document}